\def\etal{{et al.\thinspace}}

\def\spose#1{\hbox to 0pt{#1\hss}}
\def\approxlt{\mathrel{\spose{\lower 3pt\hbox{$\sim$}}
        \raise 2.0pt\hbox{$<$}}}
\def\approxgt{\mathrel{\spose{\lower 3pt\hbox{$\sim$}}
        \raise 2.0pt\hbox{$>$}}}

\def\multleft#1{\hbox to size{\vbox {\halign {\lft{##}\cr #1}}\hfill}\par}
\def\multright#1{\hbox to size{\vbox {\halign {\rt{##}\cr #1}}\hfill}\par}

\def\degmark{^\circ}
\def\boxit#1{\vbox{\hrule\hbox{\vrule\kern3pt\vbox{\kern3pt
          #1 \kern3pt}\kern3pt\vrule}\hrule}}

\def\cm{{\rm\thinspace cm}}

\def\erg{{\rm\thinspace erg}}
\def\eV{{\rm\thinspace eV}}

\def\keV{{\rm\thinspace keV}}
\def\km{{\rm\thinspace km}}

\def\Mpc{{\rm\thinspace Mpc}}

\def\ph{{\rm\thinspace ph}}
\def\s{{\rm\thinspace s}}

\def\chisq{\hbox{$\chi^2$}}

\def\pcmcu{\hbox{$\cm^{-3}\,$}}

\def\ergpcmsqps{\hbox{$\erg\cm^{-2}\s^{-1}\,$}}

\def\ergps{\hbox{$\erg\s^{-1}\,$}}

\def\kmps{\hbox{$\km\s^{-1}\,$}}

\def\pcmsq{\hbox{$\cm^{-2}\,$}}

\def\phpcmsqps{\hbox{$\ph\cm^{-2}\s^{-1}\,$}}

\def\ps{\hbox{$\s^{-1}\,$}}

\def\kmpspMpc{\hbox{$\kmps\Mpc^{-1}$}}

\documentclass[preprint]{aastex}
\usepackage{epsfig}

\received{}
\accepted{}

\slugcomment{To appear in The Astrophysical Journal} 

\shorttitle{A Chandra X-ray Study of NGC 1068}
\shortauthors{Young, Wilson \& Shopbell}

\begin{document}

\title{A Chandra X-ray Study of NGC 1068 --- I. Observations of
  Extended Emission}

\author{A. J. Young, A. S. Wilson\altaffilmark{1}}

\affil{Astronomy Department, University of Maryland, College Park, MD
  20742; ayoung@astro.umd.edu, wilson@astro.umd.edu}

\and

\author{P. L. Shopbell}

\affil{Department of Astronomy, Mail Code 105-24, California Institute
  of Technology, Pasadena, CA 91125; pls@astro.caltech.edu}


\altaffiltext{1}{Adjunct Astronomer, Space Telescope Science
  Institute, 3700 San Martin Drive, Baltimore, MD 21218;
  awilson@stsci.edu}


\begin{abstract}
  We report sub arc-second resolution X-ray imaging-spectroscopy of
  the archetypal type 2 Seyfert galaxy NGC 1068 with the Chandra X-ray
  Observatory. The observations reveal the detailed structure and
  spectra of the 13 kpc-extent nebulosity previously imaged at lower
  resolution with ROSAT. The Chandra image shows a bright, compact
  source coincident with the brightest radio and optical emission;
  this source is extended by $\simeq 1\farcs5$ (165 pc) in the same
  direction as the nuclear optical line and radio continuum emission.
  Bright X-ray emission extends $\simeq 5 \arcsec$ (550 pc) to the NE
  and coincides with the NE radio lobe and gas in the narrow line
  region. The large-scale emission shows trailing spiral arms and
  other structures. Numerous point sources associated with NGC 1068
  are seen. There is a very strong correlation between the X-ray
  emission and the high excitation ionized gas seen in HST and
  ground-based [O {\sc iii}] $\lambda 5007$ images. The X-rays to the
  NE of the nucleus are absorbed by only the Galactic column density
  and thus originate from the near side of the disk of NGC 1068. In
  contrast the X-rays to the SW are more highly absorbed and must
  come from gas in the disk or on the far side of it. This geometry is
  similar to that inferred for the narrow line region and radio lobes.
  
  Spectra have been obtained for the nucleus, the bright region
  $\simeq 4 \arcsec$ to the NE and 8 areas in the extended emission.
  The spectra are inconsistent with hot plasma models, the best
  approximations requiring implausibly low abundances ($\approxlt 0.1
  Z_\odot$). Models involving two smooth continua (either a
  bremsstrahlung plus a power-law or two bremsstrahlungs) plus
  emission lines provide excellent descriptions of the spectra. The
  emission lines cannot be uniquely identified with the present
  spectral resolution ($\sim 110$ -- 190 eV), but are consistent with
  the brighter lines seen in the XMM-Newton RGS spectrum below 2 keV.
  Hard X-ray (above 2 keV) emission, including an iron line, is seen
  extending $\simeq 20 \arcsec$ (2.2 kpc) NE and SW of the nucleus.
  Lower surface brightness, hard X-ray emission, with a tentatively
  detected iron line extends $50 \arcsec$ (5.5 kpc) to the west and
  south. Our results, when taken together with the XMM-Newton RGS
  spectrum, suggest photoionization and fluorescence of gas by
  radiation from the Seyfert nucleus to several kpc from it. The facts
  that i) the large scale (arc minute) and small scale (few arc secs)
  X-ray emissions align well and ii) the morphology of the large-scale
  emission does not correlate well with the starburst suggests that
  the starburst is not the dominant source of the extended X-rays.
\end{abstract}




%

\newpage
\section{Introduction}

Extended X-ray emission in Seyfert galaxies represents a potentially
powerful probe of these active galactic nuclei. Such X-rays could
originate from either a hot, collisionally-ionized or a much cooler
photoionized gas. Both may plausibly be expected to be present in
these objects. The narrow line regions of Seyferts are known to be the
sites of mass motions of several hundreds to $\sim 1000 \kmps$. Shocks
associated with these motions will generate gas with temperatures
$\simeq 10^6$ -- $10^7$ K. Hot gas may also be present in the form of
outflowing winds driven by radiation or radio jets from the nucleus.
The compact, hard, UV -- soft X-ray nuclear continuum source is
believed to photoionize the narrow line region. Lines from highly
ionized species, such as He {\sc ii}, [Ne {\sc v}], [Fe {\sc vii}],
[Fe {\sc x}], [Fe {\sc xi}] and [Fe {\sc xiv}], are found in the
optical and infrared spectra of these objects. X-ray lines from highly
ionized species may, therefore, also be expected. High velocity shocks
can also be powerful sources of ionizing radiation and, if present,
should provide both collisionally- and photo-ionized gas. By searching
for both of these components, X-ray observations can probe whether
shocks are significant sources of ionizing radiation for the narrow
line region. Extended X-ray emission can also arise through electron
scattering and fluorescence of the nuclear radiation in extended gas.
Morphological correspondences between the X-ray emitting gas and the
optical line and radio continuum structures may provide clues to the
nature of the X-ray emission.

For these reasons, we have begun a program of observing Seyfert
galaxies with Chandra. Previous X-ray observatories have lacked the
high spatial resolution ($\approxlt 1 \arcsec$), wide energy coverage
(0.1 -- 10 keV) and good spectral resolution ($\sim 130 \eV$ from the
CCD detectors) of Chandra. These capabilities are ideal for
investigation of extended gas in Seyferts, and in this paper we
present the results of observations of the nearby Seyfert 2 galaxy NGC
1068.

NGC 1068 was observed on three occasions with the Einstein observatory
IPC by \citet{mh87}. They found that the 0.1 -- 4.5 keV spectrum could
be described by a power law with energy index $\alpha = 2.0 \pm 0.3$
and absorbing column density consistent with the Galactic value.
Combination of EXOSAT and Einstein IPC observations \citep{el88}
showed that the X-ray spectrum can be decomposed into two components
--- a steep low energy ($< 2 \keV$) part with $\alpha = 3.5$ and a
flat high energy (2 -- 10 keV) part with $\alpha \simeq 0.7$. This was
the first detection in a Seyfert 2 of the hard power-law X-ray source
known to be present in Seyfert 1s. Any intrinsic absorption was shown
to be small ($N_H < 3 \times 10^{20} {\rm ~atom~} \pcmsq$).  Ginga
observations \citep{ketal89} confirmed the presence of the hard source
above 2 keV, and also discovered an intense iron line with equivalent
width $1.3 \pm 0.1 \keV$, in accord with the predictions of
\citet{kk87}. \citet{metal93} obtained a higher spectral resolution
(FWHM $\simeq 100 \eV$) observation with BBXRT. They found a total
equivalent width for the Fe K line of 2.8 keV, and modeled the line
profile in terms of three components corresponding to fluorescence of
neutral (and lowly ionized) iron and recombination into both He-like
and H-like iron. Fe L-shell emission was also detected. In a study
with ASCA, \citet{uetal94} found emission from the H-like
and / or He-like ions of Ne, Mg, Si and S and confirmed the Fe L and
Fe K features found by \citet{metal93}. The best fit power-law
spectrum above 3 keV has energy index $\alpha \simeq 0.3 \pm 0.3$.
\citet{uetal94} interpreted the steep low energy ($< 3 \keV$) spectrum
in terms of a two temperature (0.59 and 0.14 keV), optically-thin
coronal plasma model. \citet{ifm97} reanalyzed the ASCA data,
confirming the 3 components to the Fe K line complex which they
interpreted with a model of cold and warm reflection. \citet{nt97}
developed both photoionization and hot plasma models, including a hard
reflected continuum. Their photoionized gas model is in good agreement
with the emission lines observed by ASCA assuming solar metallicity
for all elements except iron, which is more than twice solar, and
oxygen, which is less than 0.25 solar. NGC 1068 has recently been
studied with BeppoSAX. \citet{metal97} detected the galaxy in the 20
-- 100 keV band, supporting models envisaging a mixture of both
neutral and ionized reflections of an otherwise invisible nuclear
source. The nucleus is completely obscured at all energies, implying
it is ``Compton-thick'' --- i.e. the column density of absorbing
matter exceeds $N_H \sim 10^{25} {\rm ~atoms~} \pcmsq$ (see also Matt
et al. 2000). Soft X-ray spectroscopy with BeppoSAX \citep{getal99}
reveals a spectrum rich in K$\alpha$ transitions of various elements
below 3 keV.

The only X-ray image of NGC 1068 was obtained with the ROSAT HRI by
\citet{wetal92} with $4 \arcsec$ -- $5 \arcsec$ resolution. The
observed X-ray emission could be divided into three components: i) an
unresolved (radius $\le 3 \arcsec$) source associated with the Seyfert
nucleus, ii) resolved circumnuclear (radius $< 15 \arcsec$) emission
extending preferentially towards the NE and iii) large-scale ($15
\arcsec < {\rm radius} < 60 \arcsec$) emission aligned NE-SW.
\citet{wetal92} interpreted the circumnuclear emission as thermal
emission from a nuclear wind and the large-scale emission as either
associated with the disk starburst or an extension of the nuclear wind
to larger radii. Further ROSAT HRI observations were reported by
\citet{we97}.

Very recently, \citet{petal00} have reported a high spectral
resolution observation of NGC 1068 with the XMM-Newton Reflection
Grating Spectrometer. Their spectrum reveals a large number of
emission lines below 2 keV. Based on the narrowness of the
recombination radiation continua, the fact that the forbidden lines
are stronger than the resonance lines in helium-like triplets, and the
weakness of Fe L emission compared to Fe K emission, \citet{petal00}
conclude that the emission is dominated by recombination in cool X-ray
photoionized gas.  The N and O helium-like triplets may contain a
$\approxlt 30 \%$ component in collisional equilibrium.

The present paper is organized as follows. In section
\ref{sec:obs_red} we describe the observations and their reductions.
Section \ref{sec:x_morph} provides a brief discussion of the X-ray
morphology, while section \ref{sec:x_spec} presents the X-ray spectra
obtained for the nucleus and five spatially extended regions. A
comparison of the X-ray images with those in other wavebands is
presented in section \ref{sec:comparison}, and conclusions are given
in section \ref{sec:conclusions}. Future papers will address the
compact sources in the field and present a more detailed analysis 
of the extended emission.

NGC 1068 is at a heliocentric redshift of $z=0.00379$ \citep{vetal91}
which, after correction to the frame of the microwave background and
assuming $H_0 = 50 \kmpspMpc$ and $q_0 = 0$, corresponds to 22.8 Mpc,
so $1\arcsec$ is equivalent to 110 pc.  The Galactic column density
towards NGC 1068 is $N_H ({\rm Gal}) = 2.99 \times 10^{20} \pcmsq$
\citep{m1etal96}

\section{Observations and Reduction} \label{sec:obs_red}

Since the nucleus of NGC 1068 is known to be a strong X-ray source, we
were concerned that ``pile-up'' could affect the Chandra observations.
In order to measure the Chandra count rate from the nucleus and
circumnuclear regions, we first obtained a short observation (obsid
343) of NGC 1068 on 1999 December 9 using chip S3 (backside
illuminated) of the Advanced CCD Imaging Spectrometer (ACIS;
\citeauthor{getal00} 2000).  Single exposures with a 0.1 s frame-time
were alternated with two exposures with a 0.4 s frame-time, the total
integration time being $\approxlt 1$ ks. This observation showed that
the nucleus is significantly piled up in the 0.4 s frame-time, but not
in the 0.1 s frame-time. It also demonstrated that the count rate in
the region extending $\simeq 6\arcsec$ NE from the nucleus is so high
that data obtained from this region with the default frame-time of 3.2
s would be piled up, but data obtained with a 0.4 s frame-time would
not be piled up. On this basis we decided to observe NGC 1068 for
separate, longer exposures in all of 0.1 s, 0.4 s, and 3.2 s
frame-times. The longer 0.1 s, 0.4 s and 3.2 s frame-time observations
were thus designed to study primarily the X-ray emission of the
nucleus, the extended region to the NE of the nucleus and the larger
scale emission, respectively.

These longer integrations were taken with the ACIS instrument in two
separate observations.  On 2000 February 21 the standard 3.2 s
frame-time was used. CCDs I2, I3, S1, S2, S3 and S4 were read out,
though all of the detected X-ray emission from NGC 1068 is on S3. On
2000 February 22 the readouts were alternated between one 0.1 s and
two 0.4 s frame-times in order to mitigate the effects of pile-up in
the inner regions, as discussed above. The data were screened for
times of high background count rates and aspect errors in the usual
way. The 3.2 s, 0.4 s and 0.1 s frame-time data have total good
exposure times of 47016.2 s, 11465.4 s and 1433.2 s, respectively. The
``livetimes'' for the alternating mode exposures were given incorrectly
in the header and had to be computed separately. A corrected formula
for the exposure time (G. Allen, private communication) was used, and
the resulting livetimes are equal to the frame-time multiplied by the
number of exposures taken with that frame-time.

Spectra and instrument responses were generated using {\sc ciao}
v1.1.5 and analyzed with {\sc xspec} v11.0.1. Spectra were grouped to
have at least 25 counts per energy bin to ensure that the $\chisq$
fitting of the data was valid. When extracting spectra from complex
regions the {\sc backscal} keyword (equal to the fractional area of
the chip occupied by the extraction region) was computed by hand, as
the values generated by {\sc ciao} are often incorrect.

In the absence of the gratings, X-ray spectra from CCD detectors such
as ACIS are not invertible, i.e. one cannot uniquely determine the
X-ray spectrum from the observed counts. To model such data the
following technique is used. A parameterized model of the source
spectrum is computed and folded through the instrument response. The
folded model and data are then compared using, e.g., the $\chisq$
test. Parameters of the model spectrum are adjusted to minimize the
$\chisq$ difference between the folded model and the data. The error
bars for a single parameter are given by the range over which that
parameter may be varied with $\Delta \chisq = 2.706$, keeping the
other model parameters fixed.  This is equivalent to the 90 per cent
confidence interval for a single interesting parameter.

The backside illuminated S3 chip has a high sensitivity to very soft
X-rays and significant counts at $\sim 0.1 \keV$ were seen from all
regions. However the instrument response is uncertain below 0.50 keV,
and this uncertainty increases towards lower energies. We thus
restricted our modeling to photon energies above 0.25 keV.

\subsection{Nucleus} \label{sec:reduction_nuc}

Counts were extracted from a 3\farcs75 diameter circle centered on the
nucleus. The 3.2~s frame-time data are extremely piled up and cannot
be used for spectral fitting. In the 0.4~s frame-time data there are
25251 counts corresponding to a count rate of $2.20 \pm 0.014$~ct\ps.
In the 0.1~s frame-time data there are 4760 counts corresponding to a
count rate of $3.32 \pm 0.048$~ct\ps. The difference in count rate is
due, in part, to the presence of pile-up in the 0.4~s frame-time data.
The background is negligible in both this and the NE region.

\subsection{NE Region} \label{sec:reduction_ne}

Counts were extracted from a 4\farcs1 $\times$ 4\farcs6 rectangle with
longer length in position angle (PA) = $45 \degmark$ centered on the
extended emission 4\farcs25 to the NE of the nucleus.  This region
does not overlap with the region used to extract the spectrum of the
nucleus.  In the 3.2~s frame-time data there are 26871 counts
corresponding to a count rate of $0.57 \pm 0.003$~ct\ps.  In the 0.4~s
frame-time data there are 10069 counts corresponding to a count rate
of $0.88 \pm 0.01$~ct\ps. In the 0.1 s data there are 1349 counts
corresponding to a count rate of $0.94 \pm 0.03$~ct\ps. Thus, there is
significant pile-up in the 3.2 s frame-time data in this region.

\subsection{Larger Scale Emission} \label{sec:reduction_large}

\begin{figure}
\centerline{\psfig{figure=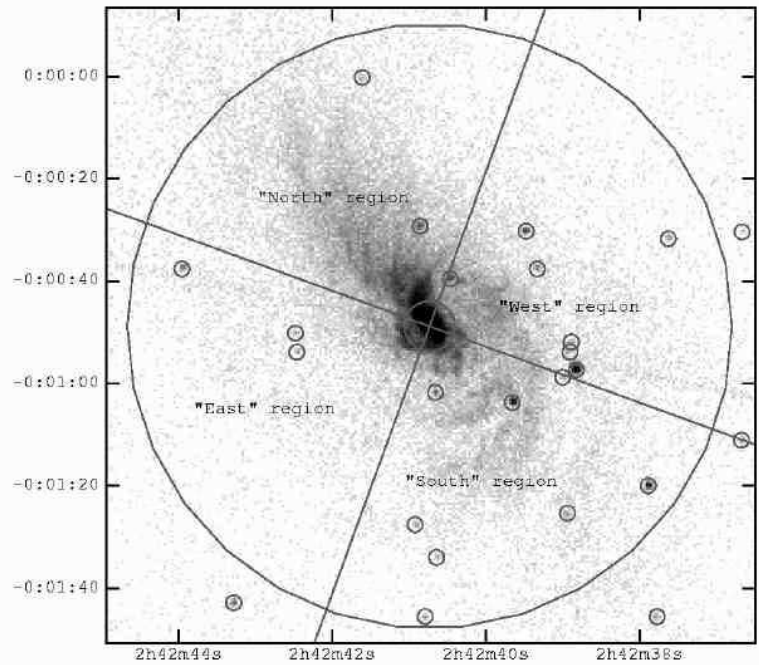,width=0.8\textwidth,angle=0}}
\caption{The extraction regions of the larger scale emission
  superposed on a grey scale image derived from the 3.2 s frame time
  data. Excluded point sources are indicated by circles.
\label{fig:extraction}}
\end{figure}

The distribution and spectrum of larger scale emission was
investigated with the 3.2 s frame-time data. The large scale region
was divided into four quadrants with position angles $250 \degmark
\rightarrow 340 \degmark$ (quadrant A), $340 \degmark \rightarrow 70
\degmark$ (quadrant B), $70 \degmark \rightarrow 160 \degmark$
(quadrant C) and $160 \degmark \rightarrow 250 \degmark$ (quadrant D).
Position angle $\sim 25 \degmark$, the bisector of quadrant B,
corresponds to the approximate position angles of the radio emission
and narrow line regions to the NE of the nucleus. Extraction regions
between nuclear distances of $5\arcsec$ and $60\arcsec$ within
quadrants A, B, C and D are termed the ``West'', ``North'', ``East''
and ``South'' regions respectively, and are shown in Figure
\ref{fig:extraction}.  Obvious point sources determined by eye were
excluded. Pile-up is not significant in any of these regions.

It is difficult to determine the spectrum of the background for the
larger scale emission as there are no source-free regions nearby. To
overcome this problem, we used a compilation of observations of
relatively blank fields, from which discrete sources have been
excised. For a particular source extraction region, a corresponding
background region is taken from this compilation. The two regions have
the same physical location on the S3 chip. The background images and
software$\footnotemark$ of Maxim Markevitch were used for this
procedure.

\footnotetext{Available at
  http://hea-www.harvard.edu/$\sim$maxim/axaf/acisbg/}

\section{X-ray Morphology} \label{sec:x_morph}

\begin{figure}
\centerline{\psfig{figure=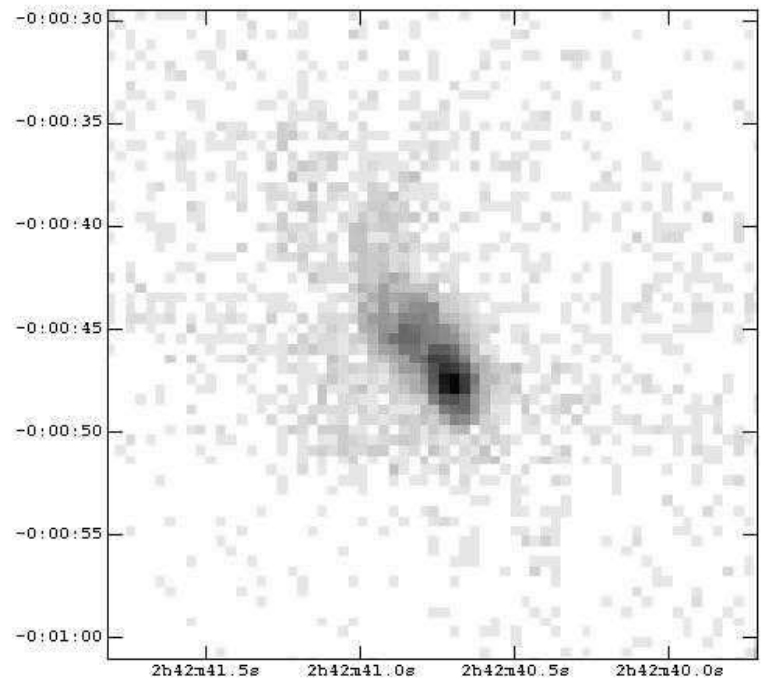,width=0.8\textwidth,angle=0}}
\caption{A grey scale representation of the Chandra X-ray
  image of NGC~1068 in the 0.25 -- 7.50 keV band taken with a 0.1 s
  frame-time. Coordinates are for epoch J2000.0, both here and in
  subsequent Figures.
\label{fig:im_01}}
\end{figure}

\begin{figure}
\centerline{\psfig{figure=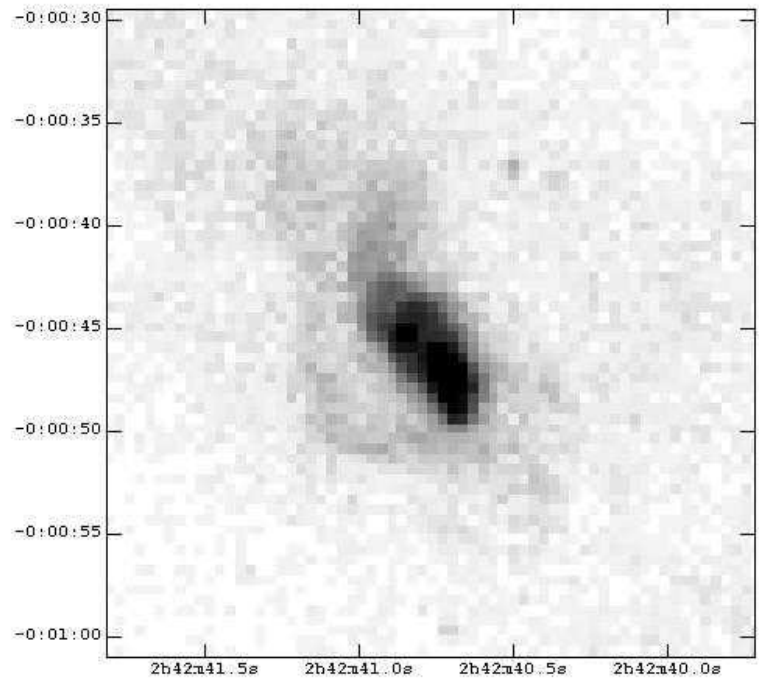,width=0.8\textwidth,angle=0}}
\caption{A grey scale representation of the Chandra X-ray
  image of NGC~1068 in the 0.25 -- 7.50 keV band taken with a 0.4 s
  frame-time.
\label{fig:im_04}}
\end{figure}

Grey scale images of the 0.1 s and 0.4 s frame-time Chandra X-ray
observations of NGC~1068 are shown in Figures \ref{fig:im_01} and
\ref{fig:im_04} respectively. The brightest region, which we
refer to as the nucleus, extends $\simeq 1\farcs5$ to the NE.
Somewhat fainter X-ray emission extends $\simeq 6\farcs5$ NE of the
nucleus in PA $\sim 35 \degmark$ -- $40 \degmark$, and perpendicular
to this direction from 2\farcs5 SE to 2\farcs5 NW of the nucleus.
There is a second peak of emission 3\farcs5 to the NE of the nucleus,
with a brightness equal to approximately 15 per cent of the nucleus;
this peak lies within the ``NE region'' (see section
\ref{sec:reduction_ne}).

\begin{figure}
\centerline{\psfig{figure=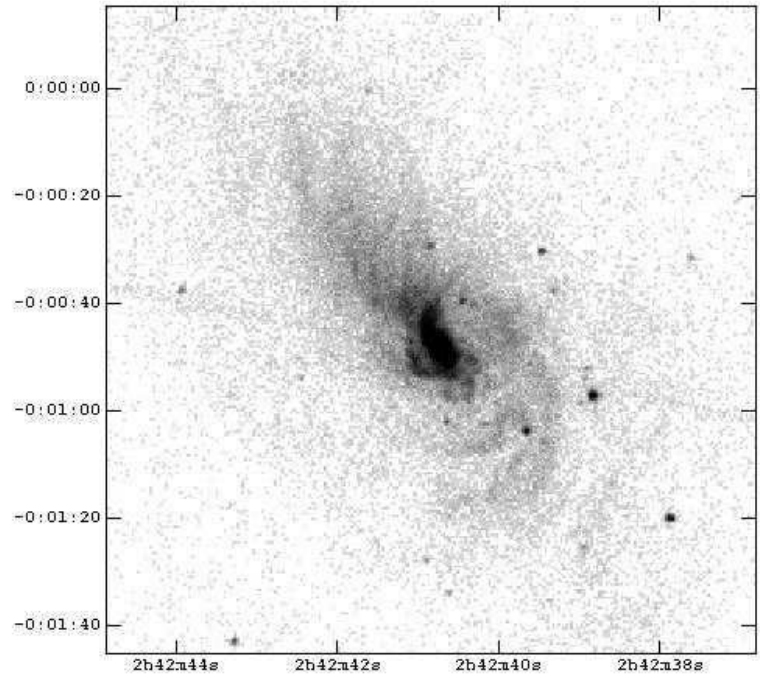,width=0.8\textwidth,angle=0}}
\caption{A grey scale representation of the Chandra X-ray
  image of NGC~1068 in the 0.25 -- 7.50 keV band taken with a 3.2 s
  frame-time. The linear feature running from PA $78\degmark$ across
  the nucleus to PA $258\degmark$ is an instrumental effect.
\label{fig:im_32}}
\end{figure}

\begin{figure}
\centerline{\psfig{figure=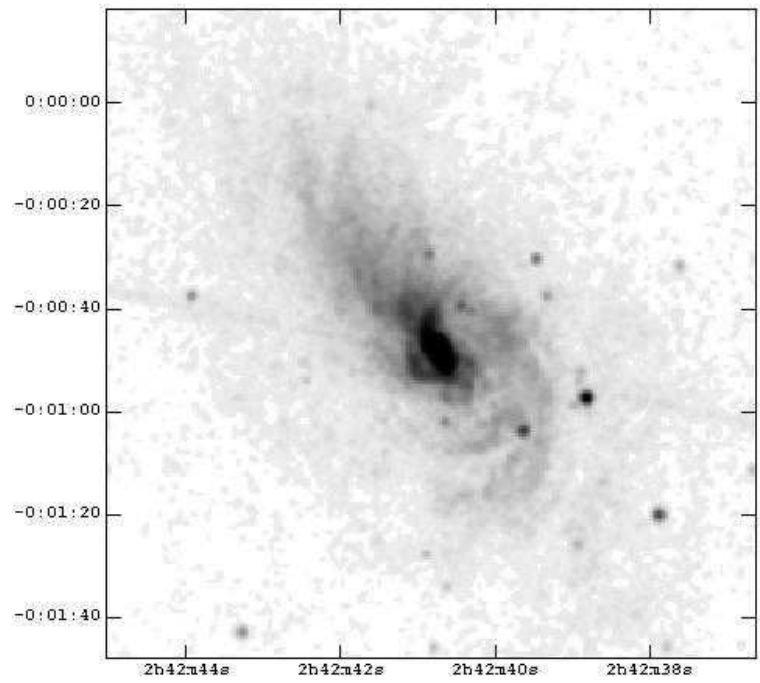,width=0.8\textwidth,angle=0}}
\caption{A grey scale representation of the Chandra X-ray
  image of NGC~1068, taken with a 3.2 s frame-time and smoothed by a
  Gaussian of $\sigma = 0\farcs5$. The linear feature running from PA
  $78\degmark$ across the nucleus to PA $258\degmark$ is an
  instrumental effect.
\label{fig:im_smoothed}}
\end{figure}

A grey scale image of the 3.2 s frame-time observation is shown in
Figure \ref{fig:im_32}. This image agrees well with the ROSAT HRI
image \citep{wetal92} when allowance is made for the lower resolution
($4 \arcsec$ -- $5 \arcsec$) of the latter. Two bright ``prongs'' of
X-ray emission extend from the bright region around the nucleus. One
extends from the northern tip along PA $330 \degmark$ for 7\farcs5,
and the second extends from the southern tip along PA $90 \degmark$
for 6\farcs5 before turning to follow PA $5 \degmark$ for 4\arcsec.
Significant extended X-ray emission is seen out to at least $60
\arcsec$ to the NE, $50 \arcsec$ to the SW, $20 \arcsec$ to the NW and
$30 \arcsec$ to the SE of the nucleus.  The most obvious large scale
structures appear to be ``spiral arms'' that curve to lower PA with
increasing radius from the nucleus, i.e.  they are ``trailing'' arms.
They may also be seen in an image of the 3.2 s frame-time data that
has been smoothed by a Gaussian of $\sigma = 0\farcs5$ (Figure
\ref{fig:im_smoothed}).  The images also show many point sources of
X-ray emission associated with NGC 1068.

\begin{figure}
\centerline{\psfig{figure=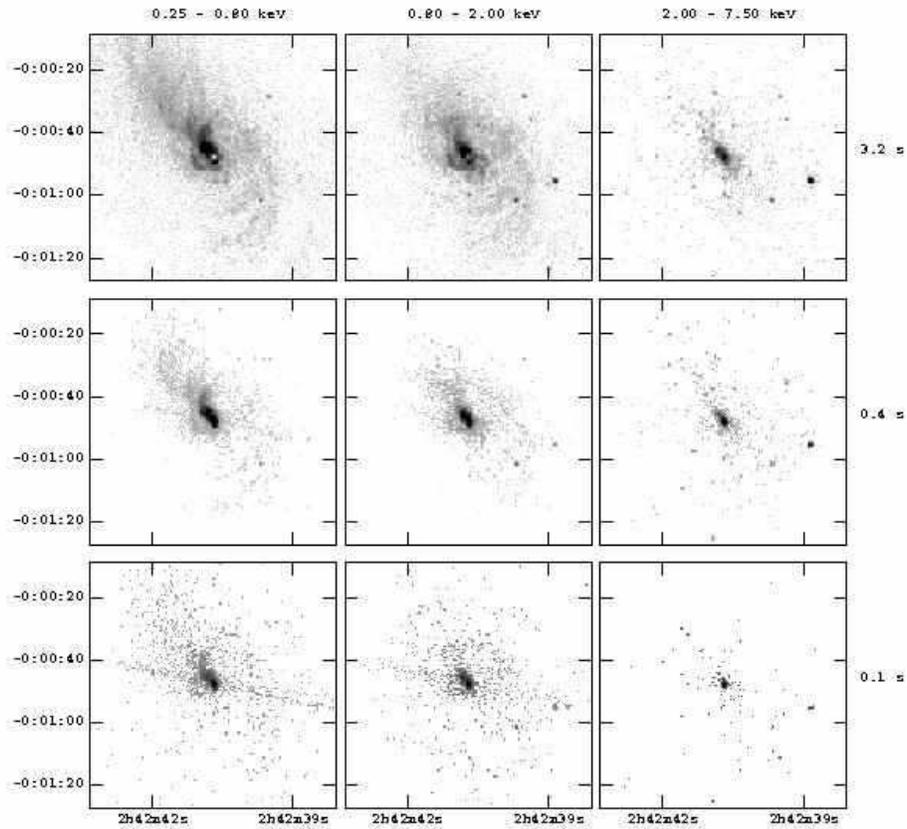,width=0.8\textwidth,angle=0}}
\caption{Grey scale representations of Chandra X-ray images
  of NGC~1068 in three energy bands and for each of the frame-times
  used.  The top row has a 3.2 s frame-time, the center row 0.4 s
  frame-time and the bottom row 0.1 s frame-time. In the left column
  the energy range is 0.25 -- 0.80 keV, the center column 0.80 -- 2.00
  keV and the right column 2.00 -- 7.50 keV.
\label{fig:im_panel}}
\end{figure}

The X-ray morphologies in the 0.25 -- 0.80 keV, 0.80 -- 2.00 keV and
2.00 -- 7.50 keV bands are shown in Figure \ref{fig:im_panel}, for the
0.1 s, 0.4 s and 3.2 s frame-time observations. The 0.25 -- 0.80 keV
X-ray emission has the greatest spatial extent, covering the range
described above. The nucleus, NE region, the two ``prongs'' and the
trailing arms are clearly seen. Only a few point sources, however,
appear in these soft X-ray images. In the 0.80 -- 2.00 keV energy
band, there is still significant emission out to tens of arc-seconds,
and many more point sources are visible. In the hardest energy band,
2.00 -- 7.50 keV, emission extends at least $\sim 13 \arcsec$ to the NE and SW
of the nucleus and at least $\sim 4 \arcsec$ to the NW and SE. The two
``prongs'' (to the N and E) do not stand out in hard X-rays, perhaps
because of the lower sensitivity in this band. As the point spread
function (PSF) at the location of the nucleus is circularly symmetric
the observed, elongated extended hard X-rays are real. In addition,
many hard point sources are clearly seen $\sim 20 \arcsec$ from the
nucleus.

\section{X-ray Spectra} \label{sec:x_spec}

\subsection{Nucleus}

\subsubsection{Thermal Plasmas}

Initially the X-ray spectrum of the nucleus was modeled as an absorbed
hot plasma with solar metal abundances.  
We used the ``{\sc mekal}''
model \citep{metal85, metal86, k92, letal95, ar85, ar92}, 
which is included in the X-ray spectral fitting software {\sc xspec}.
The 0.1 s frame-time data were used as these do not suffer from
pile-up.  The model does not fit the data (see table~\ref{tab:nuc_cont}),
so a single hot plasma may be ruled out.  Such plasmas overproduce
iron L emission lines (i.e. transitions to the $n=2$ level) between
0.7 -- 1.2 keV, and underproduce the high energy continuum above a few
keV. A better fit is obtained if the metal abundances are allowed to
vary, but the best fit value of zero is implausible for the nucleus of
a spiral galaxy. 

A model consisting of two hot plasmas, with temperatures of 0.17 keV
and 0.73 keV and metal abundances of 3.0 per cent of solar and 2.7 per
cent of solar respectively, provides an adequate description of the
spectrum below 2 keV ($\chisq = 139$ for 74 degrees of freedom
(d.o.f.)) Again, such low metal abundances are unlikely to be found in
the inner regions of a spiral galaxy, so multiple hot plasma 
models are implausible for the nucleus of NGC 1068.

\subsubsection{Bremsstrahlung Plus Power Law Plus Emission Lines}

Many lines are seen below 1 keV in the XMM-Newton RGS spectrum
(Paerels et al 2000), but these are confused by the limited spectral
resolution of the Chandra ACIS spectrum reported here. Individual
lines are sufficiently narrow to be unresolved by ACIS. We therefore
decided to take a more phenomenological approach, \label{sec:phenom}
modeling the data by a continuum plus individual emission lines.
Firstly the continuum was described by a smooth model consisting of a
bremsstrahlung plus a power law, both absorbed by the Galactic column.
The residuals between the data and this smooth model were inspected by
eye. Emission lines were then added to the model at energies expected
to reduce the value of $\chisq$. The energy and flux of these lines
were free parameters.  This process was repeated, adding lines until
either an acceptable fit was achieved, or no further improvement was
possible.  Finally, our observed line energies were compared with
those of the XMM-Newton RGS spectrum (a hard copy of which was kindly
provided by F. Paerels) and, if agreement found, identified with lines
therein.

\begin{figure}
\centerline{\psfig{figure=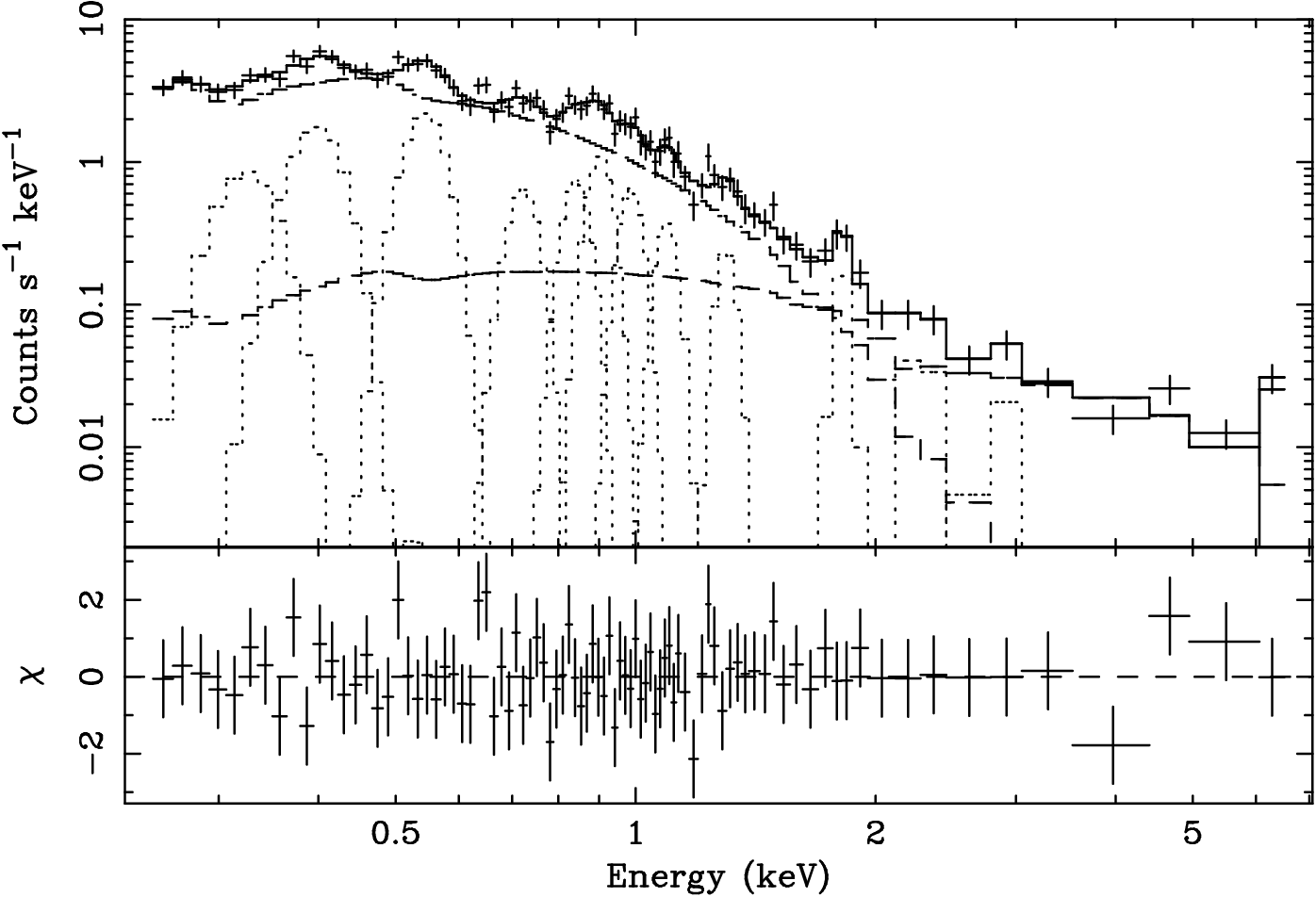,width=0.65\textwidth,angle=270}}
\caption{X-ray spectrum of the nucleus (defined in section
  \ref{sec:reduction_nuc}) of NGC~1068 extracted from the 0.1 s
  frame-time data. The upper panel shows the data points with error
  bars (crosses), with the model folded through the instrument
  response (uppermost solid line passing through the data points). The
  individual components of the model are plotted below this.  The
  lower panel shows the $\chi$ residuals from this fit.  The
  parameters of the model are listed in tables \ref{tab:nuc_cont} and
  \ref{tab:nuc_lines}. Note that the calibration is uncertain below
  0.50 keV and degrades rapidly below 0.45 keV.
\label{fig:spec_nuc}}
\end{figure}

The 0.1 s frame-time data are well described by a 0.45 keV
bremsstrahlung plasma plus a hard power law of photon index
$\Gamma=1.01^{+0.86}_{-1.15}$ plus a number of narrow emission lines,
all absorbed by the Galactic column (last model in
table~\ref{tab:nuc_cont}). The spectrum and best fitting model are
shown in Figure \ref{fig:spec_nuc} and the emission lines are listed
in table~\ref{tab:nuc_lines}. Radiative recombination continua (RRC)
are due to free-bound transitions from energy levels just above the
ionization threshold; hence these lines start at the ionization
threshold energy, and extend to higher energies (e.g. Paerels et al.
2000).

Our spectrum below 2 keV is generally consistent with lines seen in
the XMM-Newton RGS spectrum \citep{petal00}. The lines near and above
2 keV cannot be uniquely identified but are well described by neutral,
hydrogen-like or helium-like lines of Si, S or Ar, with the exception
of lines between 6.40 and 6.97 keV attributed to Fe. The Fe line
complex is well described by a neutral Fe line with an equivalent
width of 2.24 keV, although it should be noted that the true level of
the continuum at the iron line energy is difficult to determine. The
iron line is better resolved in the 0.4 s frame time data (see below).
In view of the low signal-to-noise ratio, the line flux densities are
quite uncertain.

The best fitting model has an unabsorbed 0.5 -- 2.0 keV flux of $5.8
\times 10^{-12} \ergpcmsqps$ corresponding to an unabsorbed rest frame
0.5 -- 2.0 keV luminosity of $3.6 \times 10^{41} \ergps$ and an
unabsorbed 2.0 -- 10.0 keV flux of $3.8 \times 10^{-12} \ergpcmsqps$
corresponding to an unabsorbed rest frame 2.0 -- 10.0 keV luminosity
of $2.3 \times 10^{41} \ergps$. The 2.0 -- 10.0 keV flux and
luminosity will have fractional errors comparable to the power law
component of the model, listed in table \ref{tab:nuc_cont}.

The 0.4 s frame-time observation of the nucleus suffer from pile-up
and it is difficult to determine the true shape of the continuum and
the individual line fluxes. A recent model by \citet{d01}, which is
included as part of the Interactive Spectral Interpretation System
(ISIS)\footnote{http://space.mit.edu/ASC/ISIS}, includes effects of
pile up in modeling spectra.  We fitted the same spectral model to
the 0.4 s frame-time data as was used to describe the 0.1 s frame-time
data, with the continuum parameters allowed to vary and the emission
line parameters fixed. The residuals between the data and this model
were examined by eye and any clear discrepancies noted. The fluxes of
model lines at energies near such discrepancies were then allowed to
vary in an attempt to improve the fit. If this did not provide a
satisfactory improvement, the line energy was also allowed to vary. If
no model line existed near an obvious feature a new line was added.
The reduced $\chisq$ values for these overall fits were not as good as
those for the 0.1 s frame-time data because we kept most of the line
parameters fixed, even though very small changes may have improved the
agreement between data and model.  The pile-up fraction was estimated
to be 22 per cent, and the continuum parameters ware close to those
found for the 0.1 s frame-time data; the bremsstrahlung component was
0.1 keV warmer, and the power law component was slightly harder (but
within the 0.1 s frame-time error bars) than the parameters derived
from the 0.1 s frame-time observations. These small discrepancies are
probably a result of imperfect modeling of the pile-up.

Emission-line fluxes from the 0.4 s frame-time observation are listed
in the lower portion of Table~\ref{tab:nuc_lines}. Note that the line
fluxes tabulated for the 0.4 s frame time are the new \emph{total}
fluxes of those lines that have changed or been added, and are
\emph{not} the difference in line flux between the 0.1 s and 0.4 s
frame-time observations. The 0.4 s differ from the 0.1 s results in
that: i) the O {\sc viii} RRC line is stronger (significantly above
the 0.1 s frame-time error bars), ii) an Al line at 1.72 keV is
probably present, iii) an ``absorption'' feature is seen at 1.54 keV
that coincides with a sudden change in the effective area of the
telescope, and is therefore probably caused by pile-up, iv) the Ar
line better describes the data if it is shifted to 2.97 keV, and v)
the Fe line is resolved into a neutral (6.4 keV) and ionized component
($\sim 6.80$ keV), the normalization of their sum being consistent
with the single Fe line seen in the 0.1 s frame-time data.  The fluxes
of the neutral and ionized components are comparable.

\subsection{NE Region}

The 0.4 s frame-time data do not suffer from pile-up in the NE region,
and have been used to determine both the shape of the continuum and
the emission lines present. The 3.2 s data suffer from pile-up but
have a higher signal-to-noise ratio and have been used where possible
to investigate the X-ray emission lines, especially at higher
energies.

\subsubsection{Thermal Plasmas}

The 0.4 s frame-time observations of the NE region were initially
modeled by an absorbed hot thermal plasma with solar metal
abundances, using the {\sc mekal} model in {\sc xspec}. The {\sc
  mekal} model does not provide a good description of the data (see
table~\ref{tab:ne_cont}) and may be ruled out. The model overproduces
emission lines between 0.7 and 1.2 keV due to transitions to the
L-shell of iron ions, and underproduces the high energy continuum
above a few keV. The addition of a second hot thermal plasma with
solar metal abundance improves the quality of the fit but again does
not provide a good description of the spectrum (table
\ref{tab:ne_cont}).

\subsubsection{Bremsstrahlung Plus Emission Lines}

\begin{figure}
\centerline{\psfig{figure=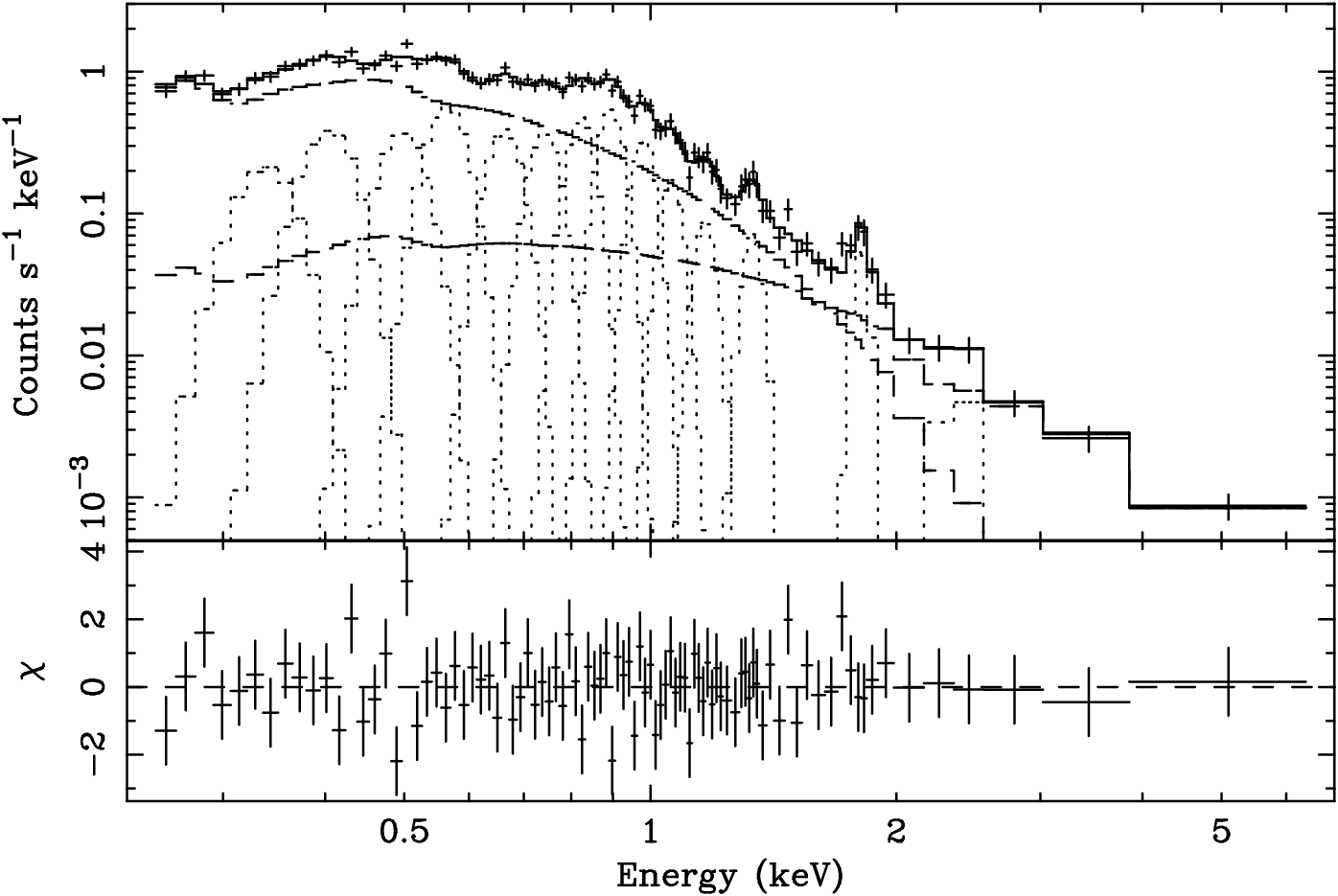,width=0.65\textwidth,angle=270}}
\caption{X-ray spectrum of the NE region (defined in
  section \ref{sec:reduction_ne}) of NGC~1068 extracted from the 0.4 s
  frame-time data. The upper panel shows the data points with error
  bars (crosses), with the model folded through the instrument
  response (uppermost solid line passing through the data points). The
  individual components of the model are plotted below this.  The
  lower panel shows the $\chi$ residuals to this fit.  The parameters
  of this fit are listed in tables \ref{tab:ne_cont} and
  \ref{tab:ne_lines}. Note that the calibration is uncertain below
  0.50 keV and degrades rapidly below 0.45 keV.
\label{fig:spec_ne}}
\end{figure}

To further investigate the emission lines in the spectrum, the
continuum was modeled by a smooth function consisting of two thermal
bremsstrahlung components absorbed by the Galactic column. No physical
meaning should be attributed to such a model, as an absorbed
bremsstrahlung plus a power law model could have been used instead.
Repeating the technique used to fit the spectrum of the nucleus,
narrow emission lines were then added to the model. The XMM-Newton RGS
spectrum (Paerels \etal 2000) was again used as a guide for the
identification of the soft X-ray lines. As noted above, the spectral
resolution of the ACIS is too low to provide unique identifications of
the soft X-ray emission lines, many of which are blended together in
our data below 1 keV. Two bremsstrahlung components with temperatures
of 0.39 keV and 2.84 keV plus the stronger emission lines seen by
\citet{petal00} and a line at 2.38 keV, all absorbed by the Galactic
column provides a good description of the spectrum with a $\chisq$ of
78 for 62 d.o.f. (bottom model, table \ref{tab:ne_cont}).  The
emission lines are listed in table~\ref{tab:ne_lines}, and the
spectrum and model are shown in Figure \ref{fig:spec_ne}.

A single data bin at 0.5 keV (see Figure \ref{fig:spec_ne})
contributes significantly to the $\chisq$ of the fit and, if removed,
the quality of fit is improved to $\chisq = 65$ for 61 d.o.f. This data
point is $9\sigma$ higher than its two neighbors over an
energy range of 15 eV (Figure \ref{fig:spec_ne}).  This is much
narrower than the instrumental response and the reason for this
feature is unknown.

The line observed in the ACIS spectrum at 0.68 keV falls in the gap
between strong lines of O {\sc viii} Ly $\alpha$, Fe {\sc xvii} and
the radiative recombination continua of N {\sc vii} and O {\sc vii}
seen in the RGS spectrum, and has been attributed to a blurred
combination of these.  The line at an observed energy of 0.76 keV is
close to O {\sc viii} Ly $\beta$ seen in the RGS spectrum, and has
been attributed to this transition. The small discrepancy in line
energy may result from blending with the RRC of O {\sc vii} at 0.74
keV. The line at 1.00 keV falls between strong lines of Fe {\sc xx}
and Ne {\sc x} seen in the RGS spectrum and may represent a
combination of these. The weak line at 2.38 keV corresponds to the
K$\alpha$ transition of sulphur in the ionization range S {\sc x} -- S
{\sc xiv}.

The best fit model to the 0.4 s frame-time data has an unabsorbed 0.5
-- 2.0 keV flux of $1.6 \times 10^{-12} \ergpcmsqps$, corresponding to
an unabsorbed rest frame 0.5 -- 2.0 keV luminosity of $1.0 \times
10^{41} \ergps$, and an unabsorbed 2.0 -- 10.0 keV flux of $2.0 \times
10^{-13} \ergpcmsqps$, corresponding to an unabsorbed rest frame 2.0
-- 10.0 keV luminosity of $1.2 \times 10^{40} \ergps$.

The 3.2 s frame-time data suffer from pile-up so we followed the same
procedure as was used for the 0.4 s frame-time observation of the
nucleus. The pile-up fraction was estimated to be 46 per cent, but
this is uncertain since the extraction region is not a point-source,
as was assumed in modeling the pile-up. The continuum parameters are
different to those found for the 0.4 s frame data; the cooler
bremsstrahlung component is 0.1 keV warmer, and the hotter
bremsstrahlung is 6.5 keV warmer. The significantly higher temperature
of the latter component may be a result of the larger energy range of
the 3.2 s frame-time data --- i.e. the continuum model is required to
produce significant X-ray flux at higher energies. The emission line
fluxes from the 3.2 s frame-time data are listed at the bottom of
Table~\ref{tab:ne_lines}.  Again, note that the tabulated line fluxes
are the \emph{total} line fluxes and not the differences in flux
between the 3.2 s and 0.4 s frame-time data. The changes are as
follows: i) the N {\sc vii} line has a preferred energy of 0.54 keV
which corresponds to a sudden change in the effective area of the
telescope, and is therefore probably caused by pile-up, ii) the Ne
{\sc ix} triplet is stronger, iii) the Ne {\sc ix} He $\beta$ line is
weaker, and iv) an Fe line is tentatively detected, but its energy is
poorly constrained,

\subsection{Larger Scale Emission}

\subsubsection{Thermal Plasmas}

We followed a similar procedure to that used for the nucleus and NE
region, applying it to the West, North, East and South regions in
turn. The 3.2 s frame-time observations were used as they have the
longest exposure time and do not suffer from pile-up.  A hot plasma
absorbed by the Galactic column does not provide a good description of
the data from any of these regions, even if the metal abundances are
allowed to vary (see table \ref{tab:extended_cont}). The preferred
values of the metal abundances are very low (less than 12 per cent
solar), and the thermal model does not fit the significant hard X-ray
emission seen from each region above 2 keV. We conclude that a single
temperature thermal plasma model, absorbed by the Galactic column,
cannot describe the spectrum in any of the larger scale extended
regions.

\subsubsection{Bremsstrahlung Plus Power Law Plus Emission Lines}

\begin{figure}
\centerline{\psfig{figure=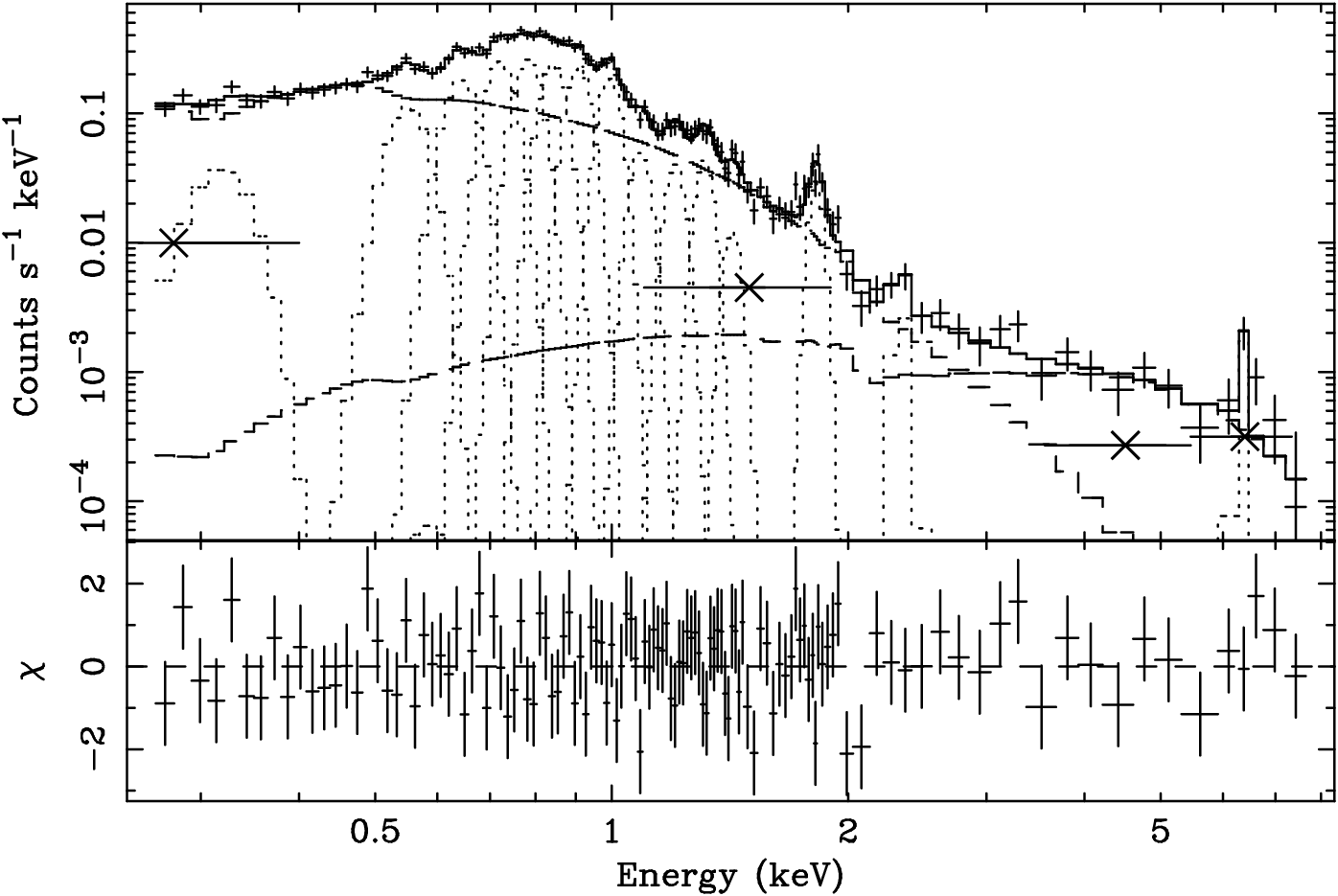,width=0.65\textwidth,angle=270}}
\caption{X-ray spectrum of the West region (defined in
  section \ref{sec:reduction_large}) of NGC~1068 extracted from the
  3.2 s frame-time data. The upper panel shows the data points with
  error bars (crosses), with the model folded through the instrument
  response (uppermost solid line passing through the data points). The
  individual components of the model are plotted below this.  The
  lower panel shows the $\chi$ residuals to this fit.  The parameters
  of this fit are listed in tables \ref{tab:extended_cont} and
  \ref{tab:west_lines}. Note that the calibration is uncertain below
  0.50 keV and degrades rapidly below 0.45 keV. The four large
  diagonal crosses with horizontal bars represent estimates of the
  contributions in the indicated energy bands from the bright, inner
  ($r < 5 \arcsec$) emission which has been scattered by the telescope
  PSF (see section \ref{sec:psf}).
\label{fig:west}}
\end{figure}

\begin{figure}
\centerline{\psfig{figure=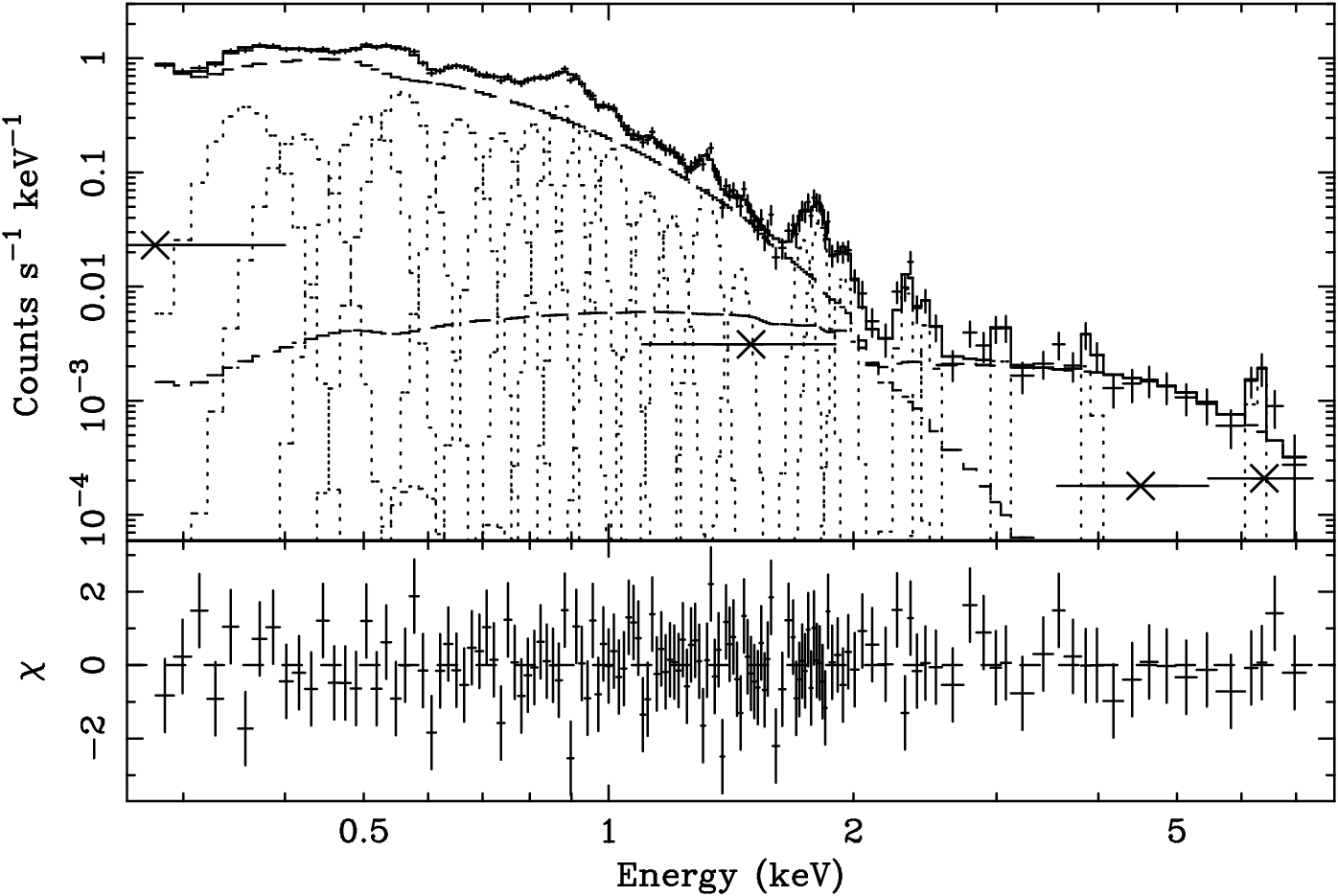,width=0.65\textwidth,angle=270}}
\caption{X-ray spectrum of the North region (defined in
  section \ref{sec:reduction_large}) of NGC~1068 extracted from the
  3.2 s frame-time data. The upper panel shows the data points with
  error bars (crosses), with the model folded through the instrument
  response (uppermost solid line passing through the data points). The
  individual components of the model are plotted below this.  The
  lower panel shows the $\chi$ residuals to this fit.  The parameters
  of this fit are listed in tables \ref{tab:extended_cont} and
  \ref{tab:north_lines}. Note that the calibration is uncertain below
  0.50 keV and degrades rapidly below 0.45 keV. The four large
  diagonal crosses with horizontal bars represent estimates of the
  contributions in the indicated energy bands from the bright, inner
  ($r < 5 \arcsec$) emission which has been scattered by the telescope
  PSF (see section \ref{sec:psf}).
\label{fig:north}}
\end{figure}

\begin{figure}
\centerline{\psfig{figure=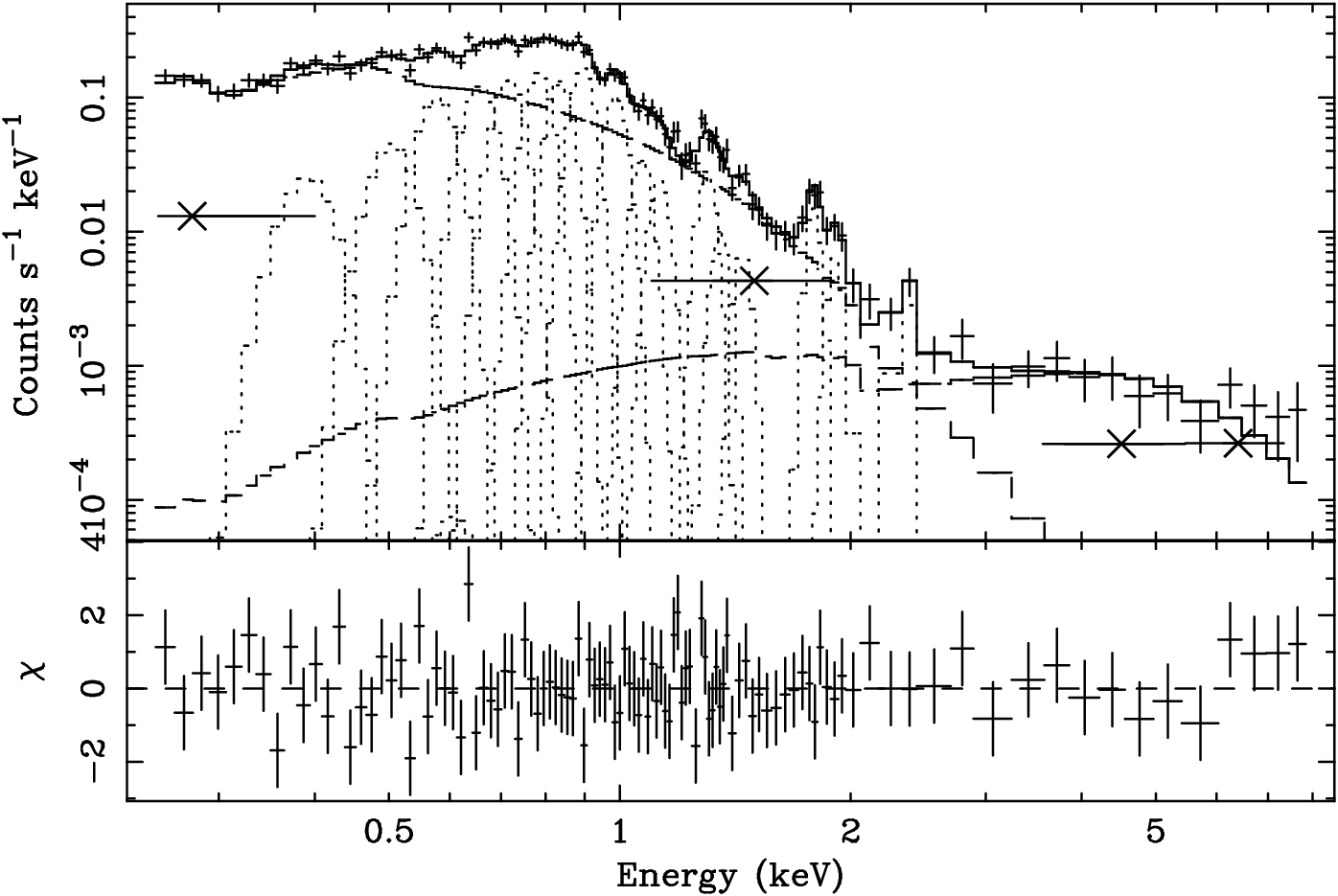,width=0.65\textwidth,angle=270}}
\caption{X-ray spectrum of the East region (defined in
  section \ref{sec:reduction_large}) of NGC~1068 extracted from the
  3.2 s frame-time data. The upper panel shows the data points with
  error bars (crosses), with the model folded through the instrument
  response (uppermost solid line passing through the data points). The
  individual components of the model are plotted below this.  The
  lower panel shows the $\chi$ residuals to this fit.  The parameters
  of this fit are listed in tables \ref{tab:extended_cont} and
  \ref{tab:east_lines}. Note that the calibration is uncertain below
  0.50 keV and degrades rapidly below 0.45 keV. The four large
  diagonal crosses with horizontal bars represent estimates of the
  contributions in the indicated energy bands from the bright, inner
  ($r < 5 \arcsec$) emission which has been scattered by the telescope
  PSF (see section \ref{sec:psf}).
\label{fig:east}}
\end{figure}

The phenomenological approach described in section \ref{sec:phenom}
was again used to study the emission lines from each of these regions.
The continuum was described by a smooth model consisting of a
bremsstrahlung plus power law, both absorbed by the Galactic column,
to which individual emission lines were then added. The parameters of
the continuum fits are listed in table \ref{tab:extended_cont}, and
the emission lines listed in tables \ref{tab:west_lines} (West),
\ref{tab:north_lines} (North), \ref{tab:east_lines} (East) and
\ref{tab:south_lines} (South). The spectrum and models are shown in
Figures \ref{fig:west} (West), \ref{fig:north} (North), \ref{fig:east}
(East) and \ref{fig:south} (South). Good fits were obtained with
bremsstrahlung components with temperatures in the range 0.36 -- 1.06
keV and power laws with photon indices in the range $\Gamma = -0.21$
to $0.92$, with the harder (lower) photon indices in the South and
East regions. There is significant hard X-ray emission above 2 keV
from each region, and an iron line is seen from the West, North and
South regions. The unabsorbed broad-band fluxes and luminosities are
listed in Table~\ref{tab:extended_flux}.

\begin{figure}
\centerline{\psfig{figure=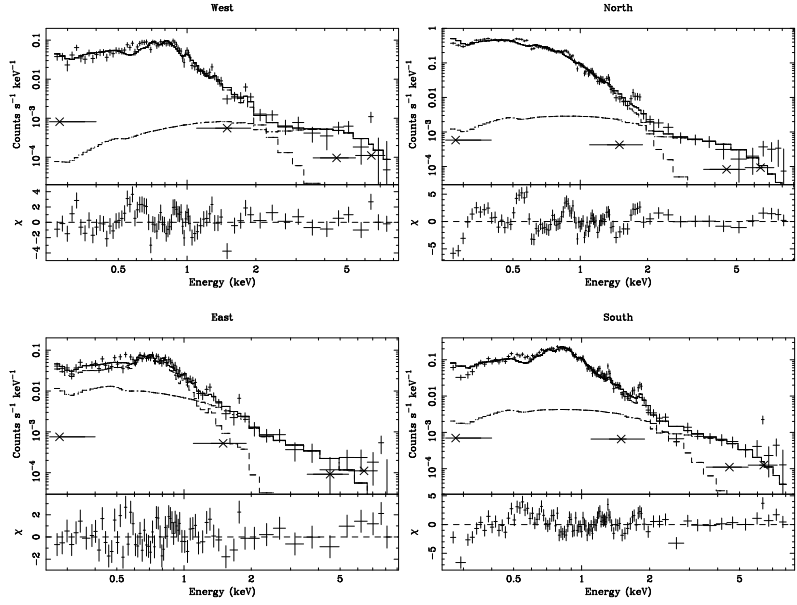,width=0.8\textwidth,angle=0}}
\caption{Spectra of the West, North, East and South
  sectors (see Figure \ref{fig:extraction}), but including only the
  annuli between radii 23\farcs5 and 60\arcsec. In each panel, the
  model is a variable metalicity {\sc mekal} plasma plus a power law.
  The four large diagonal crosses with horizontal bars represent
  estimates of the contributions in the indicated energy bands from
  the bright, inner ($r < 5 \arcsec$) emission which has been
  scattered by the telescope PSF (see section \ref{sec:psf}).
\label{fig:far_extended}}
\end{figure}

The energy dependent PSF of the Chandra mirrors will scatter a
fraction of the flux from the nucleus and NE region over tens of
arc-seconds and this can significantly contaminate the spectra of the
larger scale extended emission regions. In particular, the wings of
the PSF are stronger at higher energies and it is important to assess
the reliability of the hard X-ray emission seen on the larger scales.
Firstly, the spectra show hard ($\approxgt 3 \keV$)
X-ray fluxes differing by a factor of $\sim 3 - 4$ between the East (the
weakest) and South (the strongest) regions (see Figures \ref{fig:east}
and \ref{fig:south} respectively). In the regions between $23\farcs5$
-- $60\arcsec$ a similar difference in hard X-ray flux is also seen
(Figure \ref{fig:far_extended}), with stronger emission from the West
and South. As the PSF is essentially circularly symmetric these must
be due to differences in the intrinsic spectra of these regions.
Secondly, images of the extended hard X-ray emission (e.g. Figure
\ref{fig:im_panel}), are clearly not circularly symmetric with the
strongest emission in the large-scale regions being seen in the South
and West sectors.. This strongly suggests that scattering of nuclear
emission by the circularly symmetric PSF cannot account for all of
this emission. An ``iron line'' image was constructed by taking the
6.3 -- 6.5 keV iron line band and subtracting normalized 5.8 -- 6.3
and 6.5 -- 7.0 keV continuum band images. The resulting distribution
of discrete counts from the iron line extends $\sim 20 \arcsec$ to the
NE and SW, and at lower surface brightness $\sim 50 \arcsec$ to the
West and South. This agrees well with the hard X-ray flux and iron
lines seen in the spectra of the extended and far-extended regions
(Figures \ref{fig:west} -- \ref{fig:south} and Figure
\ref{fig:far_extended}).

To quantify the effects of scattering by the PSF \label{sec:psf} the
following approach was used. Four images were constructed in which the
central circular region of radius $5 \arcsec$ contained the observed
distribution of emission in the bands 0.25 -- 0.40, 1.10 -- 1.90, 3.56
-- 5.46 and 5.46 -- 7.34 keV. Outside this radius (which corresponds
to the inner border of the large-scale regions), the emission was set
to zero. These images were then convolved with PSFs computed at 0.23,
1.50, 3.51 and 6.40 keV respectively, using the {\sc ciao} tool {\sc
  mkarf}. From the convolved images, the predicted count rate per keV
due to telescope scattering from the inner bright region may be
obtained for each of the larger scale extended regions. These are the
four horizontal bars plotted in Figures \ref{fig:west} to
\ref{fig:far_extended}. There is clear evidence of an excess of hard
X-ray emission from the North and South regions, and tentative
evidence from the West, whereas the hard X-ray flux from the East
region may have been scattered from the nucleus by the telescope
mirrors.  Restricting our attention to the more distant regions
between $23\farcs5$ -- $60\arcsec$ (Figure \ref{fig:far_extended}) we
find there to be evidence of excess hard X-ray emission from the West
and South regions, and possibly the North.

\subsection{The Spatial Distribution of the Absorbing Column}
\label{sec:column}

\begin{figure}
\centerline{\psfig{figure=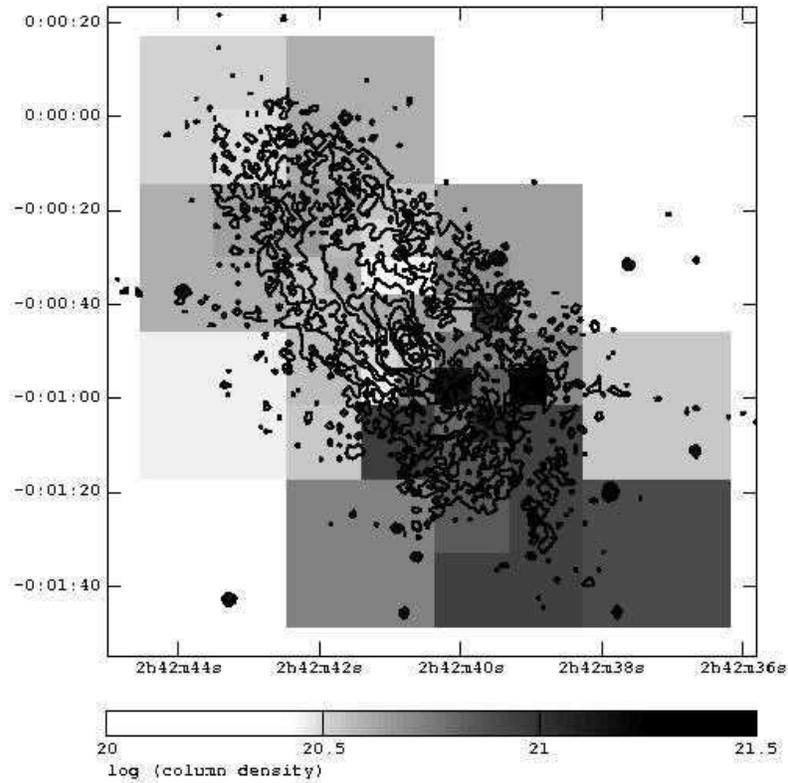,width=0.8\textwidth,angle=0}}
\caption{Contours of the 0.25 -- 7.50 keV X-ray emission
  from the 3.2 s frame-time observation superposed on a map of the
  inferred absorbing column density to the soft X-ray emission (grey
  scale). Contours are plotted at 1, 2, 4, 8, 16, 32, 64, 128, 256 and
  512 counts per pixel. The lowest contour levels are close to the
  noise, so counts have been averaged in blocks of $2 \times 2$
  pixels. The column density is seen to be larger to the SW of the
  nucleus than to the NE.
\label{fig:column}}
\end{figure}

Spectra containing at least 800 counts were extracted from square
regions of the 3.2 s frame-time Chandra data, allowing the absorbing
column density to each region to be determined. Each spectrum was
modeled by an absorbed {\sc mekal} plasma with variable metal
abundance over the energy range 0.25 -- 3.00 keV. The resulting map of
absorbing column density over the face of the galaxy is shown in
Figure \ref{fig:column}. The SW side of the galaxy shows absorption in
excess of the Galactic column ($N_H({\rm Gal})=2.99 \times 10^{20}
\pcmsq$), whereas the NE side shows no significant excess absorption.
This result shows that the extended X-ray emission to the NE of the
nucleus is on the near side of the galactic disk of NGC 1068, while
that to the SW is either in the galactic disk or on its far side.
This geometrical arrangement agrees with indications from both the
high excitation ionized gas seen at optical wavelengths and radio
continuum observations. The ionized gas to the NE is much stronger
than that to the SW, a difference usually attributed to excess
obscuration on the SW side (e.g.  \citeauthor{bww87} 1987;
\citeauthor{eetal91} 1991).  Further, the NE radio lobe is strongly
polarized while that to the SW is unpolarized (Wilson \& Ulvestad 1983,
hereafter WU),
consistent with the SW lobe being depolarized by gas in the galaxy
disk. Lastly, H {\sc i} 21 cm absorption observations indicate
unambiguously that the NE radio lobe is located in front of the galaxy
disk and the SW radio lobe is behind it \citep{getal94}.

\section{Comparison with Observations in Other Wavebands}
\label{sec:comparison}

\subsection{Registration of Images} \label{sec:registration}

NGC 1068 contains considerable structure within the central arc
second.  There are several radio components arranged in a ``bent
linear'' structure (e.g. WU; Ulvestad, Neff \& Wilson 1987; Muxlow et
al. 1996; Gallimore et al.  1996, Gallimore, Baum \& O'Dea 1997; Roy
et al. 1998). There is also structure in optical emission line and
continuum images obtained with HST (e.g. Evans et al. 1991; Macchetto
et al. 1994), and in both near- and mid-infrared band images obtained
at the diffraction limit of large ground-based telescopes (e.g.
Weinberger, Neugebauer \& Matthews 1999; Bock et al. 2000).

The location of the true nucleus (meaning the putative black hole and
its accretion disk) has been much discussed.  Radio source S1 has a
flat or inverted spectrum (Gallimore et al. 1996), which may originate
through thermal emission from the inner edge of the obscuring torus
(Gallimore, Baum \& O'Dea 1996; Roy et al. 1998).  VLBA imaging
(Gallimore, Baum \& O'Dea 1997) shows that this source is a
parsec-sized disk of ionized gas. Source S1 is also the location of
water maser emission, indicative of dense molecular gas (Claussen \&
Lo 1986; Greenhill et al. 1996). To within measurement errors, radio
source S1 is coincident with the peaks of mid-infrared (Braatz et al.
1993) and near-infrared (Thatte et al. 1997) light, and probably with
the center of symmetry of the polarization pattern of scattered UV
light (Capetti et al. 1995; Kishimoto 1999). There is thus a general
consensus that these features mark the location of the true nucleus.

There is, however, no particular reason why the peak of X-ray emission should
coincide with this location. The column density, N$_{\rm H}$, to the
X-ray nucleus of NGC 1068 exceeds $\sim$ 10$^{25}$ cm$^{-2}$ (e.g.
Matt et al. 2000), so all of the X-ray emission we observe must be
either intrinsically extended or scattered nuclear light. The Chandra
astrometry is not currently accurate enough for a precise location of
the nucleus with respect to the sub arc-second scale features
described in the previous paragraph. The position of the peak of the broad-band
X-ray emission, as measured by the 0.4s frame-time exposure, is:
$\alpha_{\rm x}$(J2000) = 02$^{h}$ 42$^{m}$ 40\fs71, $\delta_{\rm
  x}$(J2000) = --00$^{\circ}$ 00$^{\prime}$ 47\farcs7.  This position
is remarkably close to that of the brightest radio source in the
nucleus at 5 GHz (Muxlow et al. 1996): $\alpha_{\rm r}$(J2000) =
02$^{h}$ 42$^{m}$ 40\fs715s, $\delta_{\rm r}$(J2000) = --00$^{\circ}$
00$^{\prime}$ 47\farcs64.  The difference in each coordinate is $<$
0\farcs1.  This agreement is considerably better than expected given
Chandra's current absolute astrometric errors and may thus be
coincidental. Nevertheless, we decided to align the peak in the
Chandra image with the peak in a 5 GHz radio map with 0\farcs4
resolution (WU), which is similar to, though somewhat better than, the
resolution of the Chandra image.  Simulations showed that smoothing
the 60 mas resolution 5 GHz map of Muxlow et al. (1996) to the
0\farcs4 resolution of the WU map moves the radio peak slightly to the
NE (by $\le$ 0\farcs1 in each coordinate) to: $\alpha_{\rm rs}$(J2000)
= 02$^{h}$ 42$^{m}$ 40\fs721, $\delta_{\rm rs}$(J2000) =
--00$^{\circ}$ 00$^{\prime}$ 47\farcs55.  This position agrees to
$\le$ 0\farcs1 with the actual position of the peak in the WU map
after precessing the original B1950.0 coordinates of that map to
J2000.0 using the facility in the {\sc ds9} display program.

Thus, for purposes of comparison with observations in other wavebands,
the peak in the broad-band Chandra image was assigned the coordinates
$\alpha_{\rm rs}$, $\delta_{\rm rs}$ given above. In order to extend
the astrometric comparison to optical images, we used the alignment
between the optical and radio emissions obtained by Capetti, Macchetto
\& Lattanzi (1997).  These authors obtained absolute astrometry for
the HST images by taking three images: a photographic refractor plate,
a ground-based CCD image and an HST image. The plate covered
astrometric standard stars from which astrometric positions were
obtained for many fainter, secondary astrometric reference stars some
of which are on the CCD image. Finally, the astrometric scale of the
HST image was registered to the ground-based CCD image by a
cross-correlation technique. From this procedure, the location of the
peak of emission in the HST optical continuum (filter F547M) image is
found to be: $\alpha_{\rm o}$(J2000) = 02$^{h}$ 42$^{m}$ 40\fs711,
$\delta_{\rm o}$(J2000) = --00$^{\circ}$ 00$^{\prime}$ 47\farcs81
(Capetti, Macchetto \& Lattanzi 1997). This position is also very
close to the Chandra-measured location of the X-ray peak: $\alpha_{\rm
  x}$ -- $\alpha_{\rm o}$ = 0\farcs015, $\delta_{\rm x}$ --
$\delta_{\rm o}$ = 0\farcs15, which represents agreement given the
Chandra errors.

We emphasize that this astrometric registration between the X-ray
image on the one hand and the radio and optical images on the other is
not precise and may change as the Chandra astrometry improves.
However, this uncertainty only affects the alignment on sub arc-second
scales, such as is needed to compare the Chandra image with HST and
radio interferometric data. Spatial comparisons on larger scales (i.e.
$>$ 1$^{\prime\prime}$) are robust.  In the following, we compare the
X-ray morphology with that in the optical line, optical continuum and
radio continuum.

\subsection{Comparison with Optical Emission-Line Images}

\begin{figure}
\centerline{\psfig{figure=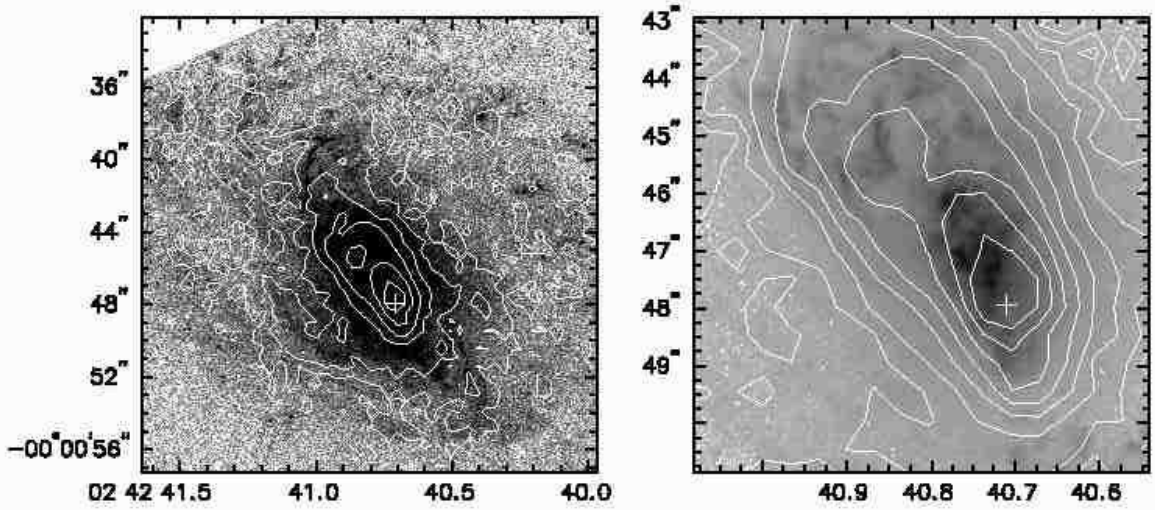,width=0.8\textwidth,angle=270}}
\caption{A superposition of the circumnuclear distribution
  of X-rays on an HST image taken through filter F502N (Dressel et al.
  1997; Capetti, Axon \& Macchetto 1997); this image is dominated by
  the light of [O {\sc iii}] $\lambda$5007. The images were aligned as
  described in Section \ref{sec:registration}. The X-ray image
  (contours) represents photon events in the energy range 0.25 -- 7.5
  keV taken from the 0.4s frame-time observation at the full Chandra
  resolution.  The cross marks the position of radio source S1, which
  is believed to coincide with the nucleus (Section
  \ref{sec:registration}). The right panel is an enlarged view of the
  circumnuclear region. {\it Left panel}: Contours are plotted on a
  logarithmic scale at log (cts pixel$^{-1}$) = 0.5, 1.0, 1.5, 2.0,
  2.5 and 3.0. The greyscale is proportional to the square root of the
  intensity and ranges from 1 $\times$ 10$^{-18}$ (white) to 1
  $\times$ 10$^{-16}$ (black) erg cm$^{-2}$ s$^{-1}$ (PC
  pixel)$^{-1}$.  {\it Right panel}: Contours are plotted on a
  logarithmic scale at log (cts pixel$^{-1}$) = 0.3, 0.6, 0.9, 1.2,
  1.5, 1.8, 2.1, 2.4, 2.7, 3.0.  The greyscale is proportional to the
  logarithm of the intensity and ranges from 1 $\times$ 10$^{-17}$
  (white) to 1 $\times$ 10$^{-13}$ (black) erg cm$^{-2}$ s$^{-1}$ (PC
  pixel)$^{-1}$. A close association is seen between many structures
  in the two wavebands.
\label{fig:hst_f502n}}
\end{figure}

\begin{figure}
\centerline{\psfig{figure=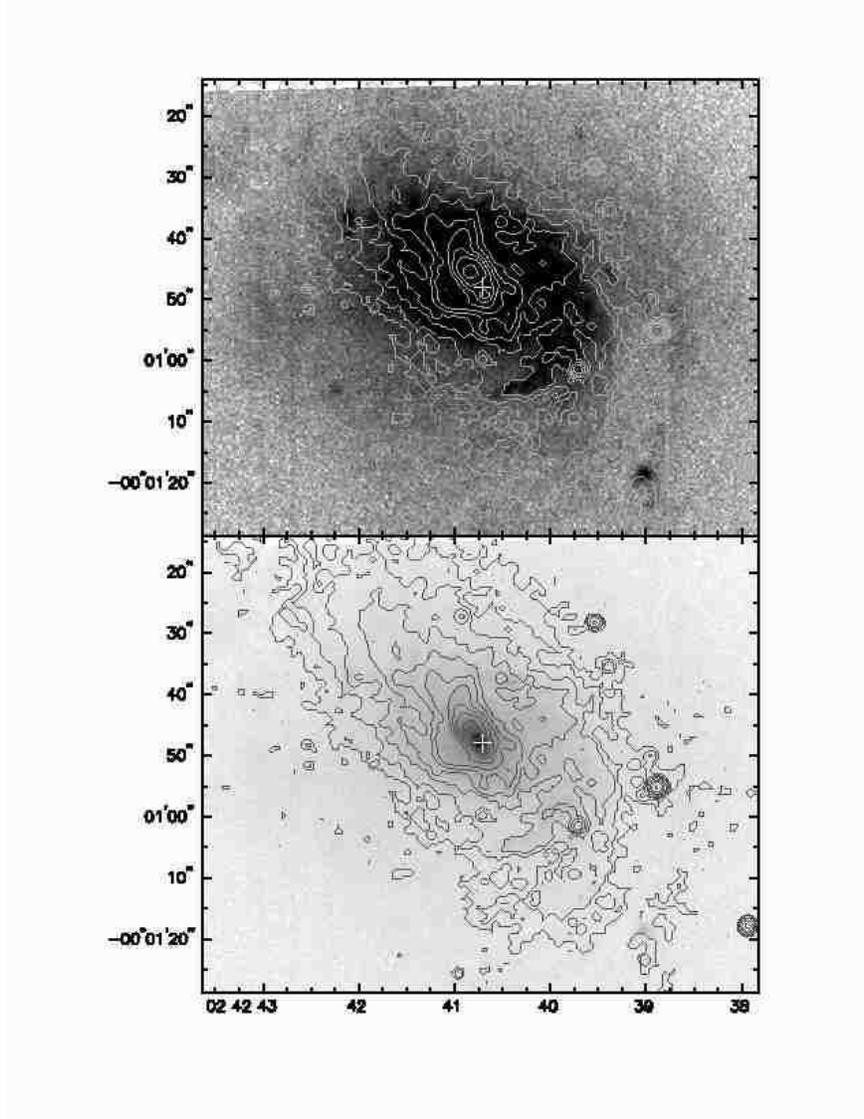,width=0.7\textwidth,angle=0}}
\caption{A superposition of the Chandra 3.2s frame-time
  data with events in the range 0.25 to 7.5 keV (contours) on an image
  in the velocity-integrated [O {\sc iii}] $\lambda$5007 line (grey
  scale) taken with the Taurus Fabry-Perot instrument on the AAT (G.
  Cecil, private communication). The images were aligned as described
  in Section \ref{sec:registration} and the cross marks the position
  of radio source S1, which is believed to coincide with the nucleus
  (Section \ref{sec:registration}).  The ``hole'' at the nucleus in
  the X-ray image is not real but results from pile-up (Section
  \ref{sec:obs_red}).  The X-ray image has been smoothed with a
  Gaussian function of standard deviation $\sigma$ = 0\farcs5.  The
  resolution (seeing) of the [O {\sc iii}] image is 1\farcs0 (FWHM),
  very similar to that of the smoothed Chandra image. The X-ray image
  is contoured at 1, 2, 4, 8, 16, 32, 64, 128, 256 and 512 cts
  (pixel)$^{-1}$ in both panels. The greyscale is proportional to the
  logarithm of the number of counts per pixel in both panels and the
  two panels differ only in the scaling of the grey scale. In the top
  panel, it ranges from log cts pixel$^{-1}$ = 0.5 (white) to 1.3
  (black) and in the bottom from log cts pixel$^{-1}$ = 0.5 (white) to
  4 (black).
\label{fig:taurus}}
\end{figure}

\begin{figure}
\centerline{\psfig{figure=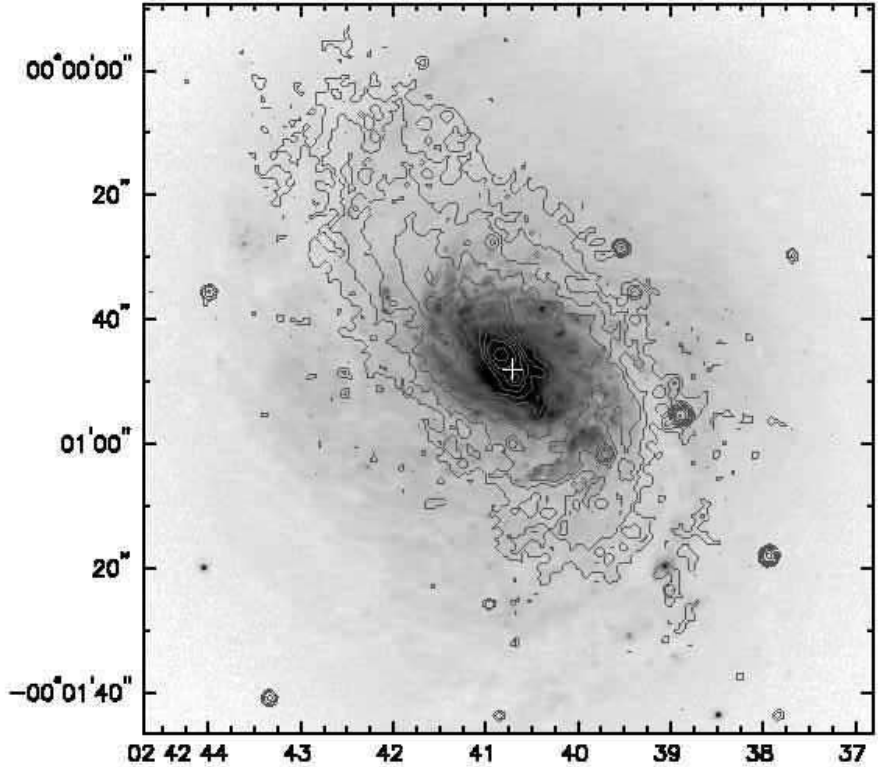,width=0.8\textwidth,angle=270}}
\caption{A superposition of the Chandra 3.2s frame-time
  data with events in the range 0.25 to 7.5 keV (contours) on an
  optical continuum image taken through a filter with center
  wavelength/bandwidth 6100/200 \AA\ (grey scale; Pogge \& De Robertis
  1993).  The images were aligned as described in Section
  \ref{sec:registration} and the cross marks the position of radio
  source S1, which is believed to coincide with the nucleus (Section
  \ref{sec:registration}).  The ``hole'' at the nucleus in the X-ray
  image is not real but results from pile-up (Section
  \ref{sec:obs_red}).  The resolution of the optical image is 0\farcs5
  -- 0\farcs6.  The Chandra image has been smoothed with a Gaussian
  function of standard deviation $\sigma$ = 0\farcs5 and is contoured
  at 1, 2, 4, 8, 16, 32, 64, 128, 256 and 512 cts pixel$^{-1}$. The
  grey scale of the optical image is proportional to the square root
  of the intensity between 1 (white) and 2000 cts pixel$^{-1}$
  (black). Noteable correlations between the two images are apparent.
\label{fig:opt}}
\end{figure}

Figure \ref{fig:hst_f502n} shows the X-ray emission (contours)
superposed on an HST image taken through filter F502N; this image is
dominated by [O {\sc iii}] $\lambda$5007.  There is remarkable
agreement between the structures in the two wavebands.  To the S of
the nucleus (assumed to be radio source S1, marked by the cross) and
within 2$^{\prime\prime}$ of it (Figure \ref{fig:hst_f502n}, right),
the structure in both wavebands is aligned approximately N-S (PA
$\simeq$ 13$^{\circ}$) and there are detailed correspondences between
the two images. To the N of the nucleus, both structures bend towards
higher PA, reaching PA $\simeq$ 35$^{\circ}$ some 4$^{\prime\prime}$
from the nucleus. It is notable that the bright extension of the
X-rays corresponds closely to the bright emission-line clouds which
extend from the nucleus to $\sim$ 2$^{\prime\prime}$ NE of it (Figure
\ref{fig:hst_f502n}, right). There is also a close correspondence
between the brightnesses slightly further (6$^{\prime\prime}$) from
the nucleus, where the X-ray contours take a V-shaped form, with apex
towards the NE (see also Figures \ref{fig:taurus} and \ref{fig:opt}).

Correspondences are also seen on the larger scale shown in Figure
\ref{fig:hst_f502n}, left. Some 7$^{\prime\prime}$ NE of the nucleus,
there is a prominent narrow feature in the HST image that first
extends N and then curves towards the NW, ending up extending in PA
$\sim$ 287$^{\circ}$ when 10$^{\prime\prime}$ from the nucleus. The
X-ray ridge, although observed with lower resolution, follows this
feature closely. An arm-like feature in the HST image begins some
4$^{\prime\prime}$ SE of the nucleus and curves towards lower PA,
extending towards the NE and ending up some 6 -- 7 $^{\prime\prime}$ E
of the nucleus; the X-ray contours follow it closely. Various other
close correspondences between the faint [O {\sc iii}] and X-ray
emission are also visible in Figure \ref{fig:hst_f502n}, left.

\begin{figure}
\centerline{\psfig{figure=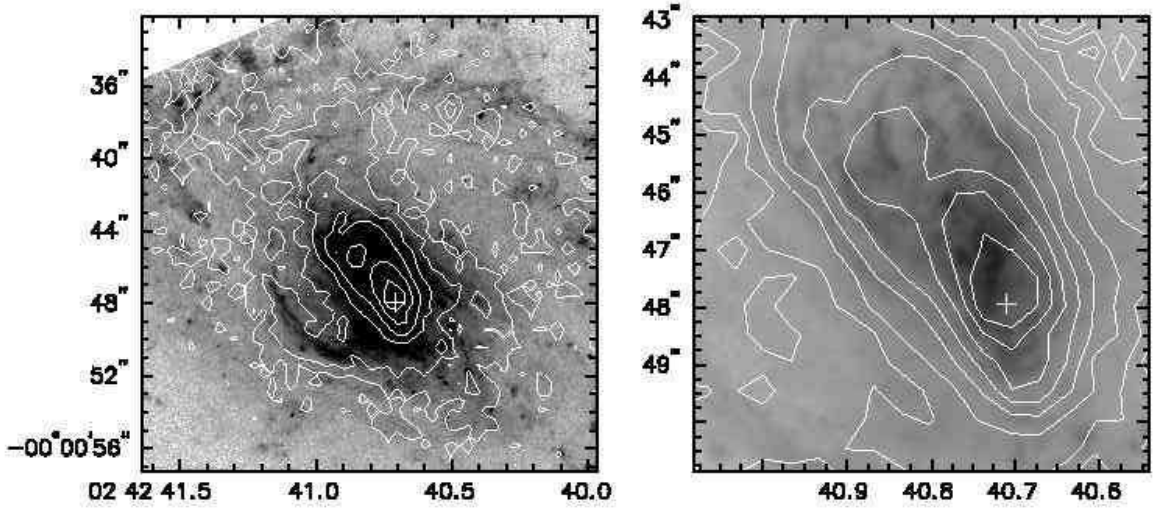,width=0.8\textwidth,angle=270}}
\caption{A superposition of the circumnuclear distribution
  of X-rays on an HST image taken through filter F658N (Dressel et al.
  1997; Capetti, Axon \& Macchetto 1997); this image is dominated by
  the light of H$\alpha$ + [N {\sc ii}] $\lambda\lambda$6548, 6583.
  The images were aligned as described in Section
  \ref{sec:registration}.  The X-ray image (contours), the contour
  levels and the cross are the same as shown in Figure
  \ref{fig:hst_f502n}. The right panel is an enlarged view of the
  circumnuclear region. {\it Left panel}: The greyscale is
  proportional to the square root of the intensity and ranges from 1
  $\times$ 10$^{-18}$ (white) to 2 $\times$ 10$^{-16}$ (black) erg
  cm$^{-2}$ s$^{-1}$ (PC pixel)$^{-1}$.  {\it Right panel}: The
  greyscale is proportional to the logarithm of the intensity and
  ranges from 1 $\times$ 10$^{-17}$ (white) to 1 $\times$ 10$^{-13}$
  (black) erg cm$^{-2}$ s$^{-1}$ (PC pixel)$^{-1}$. As for the [O {\sc
    iii}] $\lambda$5007 image (Figure \ref{fig:hst_f502n}), there are
  clear associations between structures in the two wavebands.
\label{fig:hst_f658n}}
\end{figure}

Similar correspondences are seen between the X-ray emission and the
HST image through filter F658N (Figure \ref{fig:hst_f658n}), which is
dominated by H$\alpha$ + [N {\sc ii}] $\lambda\lambda$6548, 6583.  The
relative intensities of the various line features are different in
this band (e.g. the narrow feature beginning 7$^{\prime\prime}$ NE is
much weaker here than in [O {\sc iii}]) but there is still a close
association between the structures in the two wavebands.

A comparison between [O {\sc iii}] $\lambda$5007 and X-ray emission on
a larger scale is shown in Figure \ref{fig:taurus}. The lower panel
shows again the extension from the nucleus to the NE which is common
to the two images. The overall shape of the outer X-ray contours is
similar to the [O {\sc iii}] (top panel). To the SW, there are two
spiral arms that curve outwards in a clockwise sense (i.e. they trail
the galactic rotation). The inner, brighter one is some
18$^{\prime\prime}$ and the outer, fainter one some
28$^{\prime\prime}$ from the nucleus. In both cases, there is a
remarkably precise alignment with the X-ray contours (top panel).  The
stronger, X-ray point sources are not seen in the [O {\sc iii}] image;
this is not surprising if they are X-ray binaries in NGC 1068.

\subsection{Comparison with Optical Continuum Images}

Figure \ref{fig:opt} shows the Chandra X-ray contours superposed on a
red continuum image (from Pogge \& De Robertis 1993). Features common
to both images include a protrusion 7 -- 8$^{\prime\prime}$ SW of the
nucleus and, as described in the previous subsection, the two spiral
arms SW of the nucleus and the arm-like feature SE and E of the
nucleus.  However, on the largest scales on which X-ray emission is
seen, the X-ray image is much more elongated than the optical. There
is little or no X-ray emission seen more than 30$^{\prime\prime}$ from
the nucleus in the SE or NW directions, while the X-rays extend much
further to NE and SW.  The impression gained from these comparisons
with optical data is that the X-rays correlate more strongly with the
optical line than the optical continuum emission.

This absence of a close correlation between the X-ray and optical
continuum images strongly suggests that the starburst disk is not the
dominant source of the extended X-ray emission.

\subsection{Comparison with Radio Continuum Images}

As noted in Section \ref{sec:registration}, radio continuum images
with sub arc-second resolution reveal a ``bent linear'' structure
extending over 1\farcs5.  Although the Chandra image of the central
regions (e.g. Figure  \ref{fig:hst_f502n}) does not have good enough
resolution for a meaningful comparison, the extension of the X-ray
emission immediately to the S of the nucleus is more or less N-S, like
the radio emission.  North of the nucleus, the X-ray ridge line
``bends'' towards the NE, as does the radio emission.  Capetti,
Macchetto \& Lattanzi (1997) show that this arc second-scale radio
structure lies in a region of relatively low optical line emission and
is surrounded by emission line clouds.  As argued by many authors
(e.g. Wilson \& Willis 1980; Wilson, Ward \& Haniff 1988; Evans et al.
1991; Macchetto et al. 1994; Gallimore, Baum \& O'Dea 1996), the
structure of the narrow line region implies compression of
interstellar gas by the radio ejecta.  Thus, given the close
association between the X-rays and the narrow line region (Section
5.2), the alignment between the X-rays and the arc second-scale radio
structure is expected.

\begin{figure}
\centerline{\psfig{figure=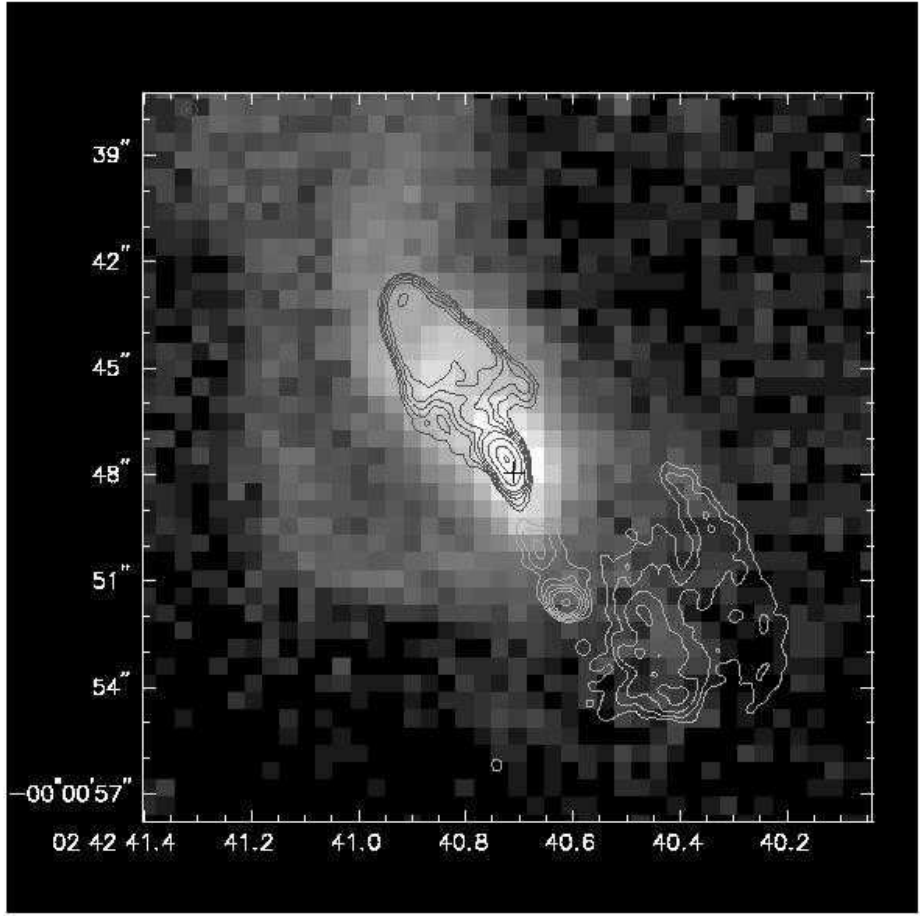,width=0.8\textwidth,angle=0}}
\caption{A superposition of the circumnuclear distribution
  of X-rays (grey scale) on a VLA 6 cm radio map (contours) with
  resolution 0\farcs38 $\times$ 0\farcs38 (Wilson \& Ulvestad 1983).
  The images were aligned as described in Section
  \ref{sec:registration}.  The X-ray image is the same as shown in
  Figure \ref{fig:im_04}. The cross marks the position of radio source
  S1, which is believed to coincide with the nucleus (Section
  \ref{sec:registration}).  Contours are plotted at 0.0005, 0.001,
  0.002, 0.004, 0.008, 0.032, 0.128 and 0.256 Jy (beam)$^{-1}$. The
  greyscale is proportional to the logarithm of the X-ray intensity
  and ranges between log cts pixel$^{-1}$ = 0 (black) and 3.0 (white).
\label{fig:vla}}
\end{figure}

A comparison between the X-ray (grey scale) and radio (contours)
images on a larger scale is shown in Figure \ref{fig:vla}. The V-shaped,
edge-brightened radio lobe to the NE, which may represent a blast wave
driven into the interstellar gas by the radio ejecta (Wilson \&
Ulvestad 1987), is seen to be associated with bright X-ray emission.
The X-ray image also exhibits a V-shaped structure with apex some
6$^{\prime\prime}$ from the nucleus (see also Figures
\ref{fig:hst_f502n}, \ref{fig:taurus}). This structure is obviously
connected with the radio ``V'', but has a wider opening angle. This
last difference may result, in part, from the lower resolution of the
X-ray image. To the SW of the nucleus, the X-rays are much weaker,
consistent with the higher absorbing column to the X-ray emitting gas
found here (Section \ref{sec:column}). Nevertheless, the brighter
regions of radio emission in the SW lobe do appear to be associated
with brighter X-ray emission. There is little or no enhancement of the
X-ray emission at the radio ``hot spot'' 4\farcs0 SW of the nucleus.

\section{Conclusions} \label{sec:conclusions}

We have obtained the highest resolution ($< 1 \arcsec$) image to date
of the X-ray emission of NGC 1068. This image shows i) a bright
nucleus close to or coincident with the brightest radio and optical
emission, ii) bright emission extending $\simeq 5 \arcsec$ (550 pc) to
the NE, iii) large-scale structure reaching at least $1 \arcmin$ (6.6
kpc) to the NE and SW, including X-ray emission from spiral arms, and
iv) numerous point sources, most of which are likely to be associated
with NGC 1068.

X-ray spectra have been obtained for ten regions -- the nucleus, the
bright region several arc-seconds to the NE, four $90 \degmark$
sectors between $5 \arcsec$ and $60 \arcsec$ from the nucleus and four
similar sectors between $23 \farcs 5$ and $60 \arcsec$ from the
nucleus. Hot plasma models are poor descriptions of the spectra and
the best approximations require implausibly low ($\approxlt 0.1
Z_\odot$) abundances. We have constructed models of the spectra
comprising two smooth continua (a bremsstrahlung plus a power-law or
two bremsstrahlung spectra) plus emission lines. The lines cannot be
uniquely identified with the present spectral resolution but are
generally consistent with the stronger lines seen in the XMM-Newton
RGS spectrum below 2 keV \citep{petal00}. Above 2 keV we observe
K$\alpha$ transitions of sulfur, argon and iron. Hard X-ray emission,
including an iron line, is seen extending $\sim 20 \arcsec$ (2.2 kpc)
NE and SW of the nucleus. Lower surface brightness hard X-ray
emission, with a tentatively detected iron line, extends $\sim 50
\arcsec$ (5.5 kpc) to the west and south. Taken together with the
XMM-Newton RGS spectrum, our results suggest photoionization and
fluorescence of gas by radiation from the Seyfert
nucleus to several kilo-parsecs from it. 

An investigation of the distribution of absorbing column shows that
the emission to the NE of the nucleus is absorbed by only the Galactic
column and is thus on the near side of the disk of NGC 1068. The X-ray
emission to the SW of the nucleus suffers greater absorption and must
originate in or behind the disk of NGC 1068. This geometry agrees with
that inferred for both the narrow line region observed at optical
wavelengths and the radio ejecta.

We have compared the X-ray image with optical emission-line, optical
continuum and radio continuum images. There is a very strong
correlation between the X-ray emission and high excitation optical
line emission (as traced by [O {\sc iii}] $\lambda 5007$), consistent
with photoionization in both wavebands. To the SW, the X-ray spiral
arms correlate closely with those seen in [O {\sc iii}] $\lambda
5007$. The NE radio lobe some $6 \arcsec$ (660 pc) from the nucleus
shows a strong morphological correlation with the X-ray emission,
suggestive of interstellar gas being compressed by the radio ejecta.
The alignment of the large-scale (arc min) and small scale (few arc
sec) X-ray emissions and the lack of a close correlation between the
larger scale X-ray emission and the optical continuum strongly
suggests that the starburst is not the dominant source of these
X-rays.

We plan to present a discussion of the compact X-ray sources seen in NGC 1068
and a more detailed analysis of the extended emission in future papers.

We thank the Chandra Science Center, especially Dan Harris and Shanil
Virani, for assistance with the observations. Glenn Allen gave
valuable advice on the alternating exposure mode. We are grateful to
Gerald Cecil and Frits Paerels for providing data in advance of
publication. We thank John Houck for assistance with ISIS. This
research was supported by NASA grant NAG 81027.

\vfil\eject\clearpage
\begin{deluxetable}{ccccccc}
\tabletypesize{\footnotesize}
\tablewidth{0pt} \rotate \tablecaption{Spectral Fits to the Nucleus
  from 0.1 s Frame-Time Observation
\label{tab:nuc_cont}}
\tablecolumns{7} \tablehead{\colhead{Model\tablenotemark{a}} &
  \colhead{$N_H$} & \colhead{Temperature} & Abundance &
  \colhead{Photon index} & \colhead{Normalization\tablenotemark{b}} &
  \colhead{$\chisq$ / d.o.f.}  \\ \colhead{} & \colhead{[$\times
    10^{20}\pcmsq$]} & \colhead{[\keV]} & \colhead{[$\times Z_\odot$]}
  & \colhead{} & \colhead{} & \colhead{}} \startdata

MEKAL & 2.99\tablenotemark{c} & 0.65 & 1.0 \tablenotemark{c} & &
$K_{\rm MEKAL} = 1.6\times 10^{-3}$ & 2415 / 88 \\

Brem & 2.99\tablenotemark{c} & 0.53 & & & $K_{\rm Brem} =
1.1\times 10^{-2}$ & 321 / 88 \\

Brem + PL & 2.99\tablenotemark{c} & 0.46 & & 0.83 & $\left\{
  \begin{array}{c} K_{\rm Brem} = 1.2\times 10^{-2} \\ K_{\rm PL} =
    2.0\times 10^{-4} \end{array} \right\}$ & 202 / 86 \\

Brem + PL + lines & 2.99\tablenotemark{c} & $0.45^{+0.03}_{-0.08}$ &
& $1.01^{+0.86}_{-1.15}$ & $\left\{ \begin{array}{c} K_{\rm
      Brem} = \left( 1.0^{+0.1}_{-0.1} \right)\times 10^{-2} \\
    K_{\rm PL} = \left( 2.3^{+3.4}_{-1.9}
    \right) \times 10^{-4} \end{array} \right\}$ & 67 / 64 \\

\enddata

\tablenotetext{a}{The model abbreviations are MEKAL = Mewe, Kaastra \&
  Liedahl thermal plasma, PL = power law, Brem = bremsstrahlung.}

\tablenotetext{b}{The model normalization are $K_{\rm MEKAL} =
  10^{-14} \int n_e n_H dV / (4 \pi ((1+z)D_A)^2)$, $K_{\rm Brem} =
  3.02 \times 10^{-15} \int n_e n_I dV / (4 \pi D^2)$ and $K_{\rm PL}
  = \phpcmsqps {\rm keV}^{-1}$ at 1 keV. $n_e$ is the electron density
  (\pcmcu), $n_H$ is the hydrogen density (\pcmcu), $n_I$ is the ion
  density (\pcmcu), $D$ is the distance to the source (cm) and $D_A$
  is the angular size distance to the source (cm).}

\tablenotetext{c}{Fixed parameter.}

\end{deluxetable}

\vfil\eject\clearpage
\begin{deluxetable}{ccccc}
\tabletypesize{\footnotesize}
\tablewidth{0pt}
\tablecaption{X-ray Spectral Lines From the Nucleus
\label{tab:nuc_lines}}
\tablecolumns{5} \tablehead{\colhead{Frame-time} & \colhead{Energy} &
  \colhead{Line} & \colhead{Observed energy} &
  \colhead{$K$\tablenotemark{a}} \\ \colhead{[sec]} & \colhead{[keV]}
  & \colhead{} & \colhead{[keV]} & \colhead{}} \startdata

0.1 & 0.37 & C {\sc vi} Ly $\alpha$ & $0.34^{+0.02}_{-0.01}$ & $\left(
  3.4^{+1.6}_{-1.9} \right) \times 10^{-4}$ \\

0.1 & 0.43 & N {\sc vi} triplet & $0.42^{+0.01}_{-0.01}$ & $\left(
  5.1^{+1.7}_{-1.9} \right) \times 10^{-4}$ \\


0.1 & 0.56 -- 0.57 & O {\sc vii} triplet & $0.56^{+0.01}_{-0.01}$ &
$\left( 4.9^{+1.8}_{-1.2} \right) \times 10^{-4}$ \\


0.1 & $\left\{ \begin{array}{c} 0.73 \\ 0.74 \end{array} \right.$ &
\begin{tabular}{c} Fe L {\sc xvii} \\ O {\sc
    vii} RRC \end{tabular} $\left. \begin{array}{c} \\ \\ \end{array}
\right\}$ & $0.74^{+0.01}_{-0.02}$ & $\left( 9.3^{+4.8}_{-5.3} \right)
\times 10^{-5}$ \\

0.1 & 0.87 & O {\sc viii} RRC & $0.86^{+0.02}_{-0.03}$ & $\left(
  9.6^{+4.1}_{-5.9} \right) \times 10^{-5}$ \\

0.1 & 0.90 -- 0.92 & Ne {\sc ix} triplet & $0.92^{+0.03}_{-0.02}$ & $\left(
  1.4^{+0.3}_{-0.6} \right) \times 10^{-4}$ \\

0.1 & 1.02 & Ne {\sc x} Ly $\alpha$ & $1.01^{+0.01}_{-0.02}$ & $\left(
  7.6^{+3.0}_{-3.1} \right) \times 10^{-5}$ \\

0.1 & 1.07 & Ne {\sc ix} He $\beta$ & $1.11^{+0.02}_{-0.02}$ & $\left(
  4.3^{+3.0}_{-3.1} \right) \times 10^{-5}$ \\

0.1 & 1.33 -- 1.34 & Mg {\sc xi} triplet & $1.31^{+0.04}_{-0.05}$ &
$\left( 2.8^{+1.7}_{-1.9} \right) \times 10^{-5}$ \\

0.1 & 1.84 -- 1.86 & Si {\sc xiii} triplet & $1.84^{+0.03}_{-0.04}$ &
$\left( 2.9^{+1.4}_{-1.4}\right) \times 10^{-5}$ \\

0.1 & 2.31 -- 2.35 & S {\sc ii} -- S {\sc x} & $2.30^{+0.03}_{-0.04}$
& $\left( 3.6^{+2.1}_{-2.1} \right) \times 10^{-5}$ \\

0.1 & 2.96 -- 3.00 & Ar {\sc ii} -- Ar {\sc xi} &
$2.85^{+0.15}_{-0.06}$ & $\left( 1.8^{+1.3}_{-1.4} \right) \times 10^{-5}$ \\

0.1 & 6.40 -- 6.97 & Fe {\sc ii} -- Fe {\sc xxvi} &
6.40\tablenotemark{b} & $\left( 7.3^{+3.3}_{-3.3} \right) \times
10^{-5}$
\\

\hline

0.4 & 0.87 & O {\sc viii} RRC & 0.86 & $2.2 \times 10^{-4}$ \\

0.4 & --- & Artifact & 1.54 & $-5.5 \times 10^{-5}$ \\

0.4 & 1.73 & Al {\sc xiii} Ly $\alpha$ & 1.72 & $2.5 \times 10^{-5}$ \\

0.4 & 2.96 -- 3.00 & Ar {\sc ii} -- Ar {\sc xi} & 2.97 & $1.8 \times
10^{-5}$ \\

0.4 & 6.40 & Fe {\sc ii} & 6.40 & $4.3 \times 10^{-5}$ \\

0.4 & $\left\{ \begin{array}{c} 6.70 \\ 6.97 \end{array} \right.$ &
\begin{tabular}{c} Fe {\sc xxv} \\ Fe {\sc xxvi} \end{tabular}
$\left. \begin{array}{c} \\ \\ \end{array} \right\}$ & 6.80 & $5.5
\times
10^{-5}$ \\

\enddata

\tablenotetext{a}{$K = {\rm total} \phpcmsqps$ in the line.}

\tablenotetext{b}{Fixed parameter.}

\end{deluxetable}

\vfil\eject\clearpage
\begin{deluxetable}{cccccc}
\tabletypesize{\footnotesize}
\tablewidth{0pt} \rotate \tablecaption{Spectral Fits to the NE Region
  from the 0.4 s Frame-Time observation
\label{tab:ne_cont}}
\tablecolumns{6} \tablehead{\colhead{Model\tablenotemark{a}} &
  \colhead{$N_H$} & \colhead{Temperature} & Abundance &
  \colhead{Normalization\tablenotemark{b}} & \colhead{$\chisq$ /
    d.o.f.} \\ \colhead{} & \colhead{[$\times 10^{20}\pcmsq$]} &
  \colhead{[keV]} & \colhead{[$\times Z_\odot$]} & \colhead{} &
  \colhead{}} \startdata

MEKAL & 2.99\tablenotemark{c} & 0.63 & 1.0 & $K_{\rm MEKAL} =
5.1\times 10^{-4}$ & 4201 / 92\\

2 $\times$ MEKAL & 2.99\tablenotemark{c} & $\left\{ \begin{array}{c}
    0.21 \\ 1.48 \end{array} \right.$ & $\begin{array}{c} 1.00 \\ 1.00
\end{array}$ & $\left. \begin{array}{c} K_{\rm MEKAL} = 7.8\times
    10^{-4} \\ K_{\rm MEKAL}  = 5.5\times 10^{-4} \end{array}
\right\}$ & 1780 / 90 \\

2 $\times$ Brem + lines & 2.99\tablenotemark{c} &
$\left\{ \begin{array}{c} 0.39^{+0.02}_{-0.05} \\ 2.84^{+0.95}_{-0.41}
    \end{array} \right.$ & & $\left.
  \begin{array}{c} K_{\rm Brem} = \left( 2.5^{+0.2}_{-0.1} \right)
    \times 10^{-3} \\ K_{\rm Brem} = \left( 1.2^{+0.8}_{-0.3} \right)
    \times 10^{-4} \end{array} \right\}$ & 78 / 62 \\

\enddata

\tablenotetext{a}{The model abbreviations are MEKAL = Mewe, Kaastra \&
  Liedahl thermal plasma, PL = power law, Brem = bremsstrahlung.}

\tablenotetext{b}{The model normalization are $K_{\rm MEKAL} =
  10^{-14} \int n_e n_H dV / (4 \pi ((1+z)D_A)^2)$, $K_{\rm Brem} =
  3.02 \times 10^{-15} \int n_e n_I dV / (4 \pi D^2)$ and $K_{\rm PL}
  = \phpcmsqps {\rm keV}^{-1}$ at 1 keV. $n_e$ is the electron density
  (\pcmcu), $n_H$ is the hydrogen density (\pcmcu), $n_I$ is the ion
  density (\pcmcu), $D$ is the distance to the source (cm) and $D_A$
  is the angular size distance to the source (cm).}

\tablenotetext{c}{Fixed parameter.}

\end{deluxetable}

\vfil\eject\clearpage
\begin{deluxetable}{ccccc}
\tabletypesize{\footnotesize}
\tablewidth{0pt}
\tablecaption{X-ray Spectral Lines From the NE Region
\label{tab:ne_lines}}
\tablecolumns{5} \tablehead{\colhead{Frame-time} & \colhead{Energy} &
  \colhead{Line} & \colhead{Observed energy} &
  \colhead{$K$\tablenotemark{a}} \\ \colhead{[sec]} & \colhead{[keV]}
  & \colhead{} & \colhead{[keV]} & \colhead{}} \startdata

0.4 & 0.37 & C {\sc vi} Ly $\alpha$ & $0.36^{+0.01}_{-0.01}$ & $\left(
  7.6^{+2.4}_{-2.5} \right) \times 10^{-5}$ \\

0.4 & 0.43 & N {\sc vi} triplet & $0.42^{+0.01}_{-0.01}$ & $\left(
  1.0^{+0.3}_{-0.3} \right) \times 10^{-4}$ \\

0.4 & 0.50 & N {\sc vii} & $0.52^{+0.00}_{-0.01}$ & $\left(
  7.1^{+1.6}_{-1.5} \right) \times 10^{-5}$ \\


0.4 & 0.59 & N {\sc vii} Ly $\beta$ & $0.58^{+0.00}_{-0.00}$ & $\left(
  1.1^{+0.2}_{-0.1} \right) \times 10^{-4}$ \\

0.4 & $\left\{ \begin{array}{c} 0.65 \\ 0.67 \\ 0.73 \\ 0.74
  \end{array} \right.$ & \begin{tabular}{c} O {\sc viii} Ly $\alpha$
  \\ N {\sc vii} RRC \\  Fe L {\sc xvii} \\ O {\sc vii}
RRC \end{tabular} $\left. \begin{array}{c} \\ \\ \\ \\ \end{array}
\right\}$ & $0.68^{+0.00}_{-0.01}$ & $\left( 6.2^{+1.1}_{-1.1} \right)
\times 10^{-5}$ \\

0.4 & 0.78 & O {\sc viii} Ly $\beta$& $0.76^{+0.01}_{-0.01}$ & $\left(
  5.4^{+1.0}_{-1.2} \right) \times 10^{-5}$ \\

0.4 & $\left\{ \begin{array}{c} 0.79 - 0.82 \\ 0.82 \end{array}
\right.$ & \begin{tabular}{c} Fe {\sc xvii} \\ O {\sc viii} Ly
  $\gamma$ \end{tabular} $\left. \begin{array}{c} \\ \\ \end{array}
\right\}$ & $0.84^{+0.00}_{-0.01}$ & $\left( 6.3^{+0.8}_{-1.2} \right)
\times 10^{-5}$ \\


0.4 & 0.90 -- 0.92 & Ne {\sc ix} triplet & $0.91^{+0.00}_{-0.01}$ &
$\left( 6.7^{+0.8}_{-0.9} \right) \times 10^{-5}$ \\

0.4 & 0.97 \& 1.02 & Fe {\sc xx} \& Ne {\sc x} Ly $\alpha$ &
$1.00^{+0.00}_{-0.01}$ & $\left( 3.9^{+0.5}_{-0.7} \right) \times
10^{-5}$ \\

0.4 & 1.07 & Ne {\sc ix} He $\beta$ & $1.08^{+0.01}_{-0.01}$ & $\left(
  2.0^{+0.5}_{-0.5} \right) \times 10^{-5}$ \\

0.4 & 1.21 & Ne {\sc x} Ly $\beta$ & $1.19^{+0.01}_{-0.02}$ & $\left(
  9.7^{+3.8}_{-3.4} \right) \times 10^{-6}$ \\

0.4 & 1.35 & Mg {\sc xi} triplet & $1.35^{+0.01}_{-0.02}$ & $\left(
  7.9^{+2.9}_{-2.9} \right) \times 10^{-6}$ \\

0.4 & 1.80 -- 1.86 & Si {\sc x} -- Si {\sc xiii} triplet &
$1.82^{+0.02}_{-0.02}$ & $\left( 7.6^{+2.2}_{-2.4} \right) \times 10^{-6}$ \\

0.4 & 2.35 -- 2.43 & S {\sc x} -- S {\sc xiv} & $2.38^{+0.04}_{-0.03}$
& $\left( 4.4^{+2.8}_{-2.8} \right) \times 10^{-6}$ \\

\hline

3.2 & $\left\{ \begin{array}{c} 0.50 \\ 0.54 \end{array} \right.$ &
\begin{tabular}{c} N {\sc vii} \\ Artifact \end{tabular}
$\left. \begin{array}{c} \\ \\ \end{array} \right\}$ & 0.54 & $1.1
\times 10^{-4}$ \\

3.2 & 0.90 -- 0.92 & Ne {\sc ix} triplet & 0.91 & $1.2 \times 10^{-4}$
\\

3.2 & 1.07 & Ne {\sc ix} He $\beta$ & 1.08 & $1.1 \times 10^{-5}$ \\



3.2 & 6.40 & Fe {\sc ii} & 6.40 & $8.1 \times 10^{-6}$ \\

\enddata

\tablenotetext{a}{$K = {\rm total} \phpcmsqps$ in the
  line.}

\tablenotetext{b}{Fixed parameter.}

\end{deluxetable}

\vfil\eject\clearpage
\begin{deluxetable}{cccccccc}
\tabletypesize{\footnotesize}
\tablewidth{0pt}
\rotate
\tablecaption{Spectral Fits to the Larger Scale Extended X-ray Emission
\label{tab:extended_cont}}
\tablecolumns{8} \tablehead{\colhead{Region} &
  \colhead{Model\tablenotemark{a}} & \colhead{$N_H$} &
  \colhead{Abundance} & \colhead{Temperature} & \colhead{Photon index}
  & \colhead{Normalization\tablenotemark{b}} & \colhead{$\chisq$ /
    d.o.f.}\\ \colhead{} & \colhead{} & \colhead{[$\times
    10^{20}\pcmsq$]} & \colhead{[$\times Z_\odot$]} & \colhead{[keV]}
  & \colhead{}}

\startdata

West & MEKAL & 2.99\tablenotemark{c} & 0.12 & 0.63 & & $K_{\rm MEKAL}
= 1.1 \times 10^{-3}$ & 481 / 122 \\

West & Brem + PL + lines & 2.99\tablenotemark{c} & &
$0.73^{+0.04}_{-0.04}$ & $0.13^{+0.63}_{-0.55}$ & $\left\{
  \begin{array}{c} K_{\rm Brem} = \left( 3.4^{+0.3}_{-0.1}
\right) \times 10^{-4} \\ K_{\rm PL} = \left( 2.9^{+8.5}_{-0.6}
\right) \times 10^{-6} \end{array} \right\}$ & 102 / 89 \\

North & MEKAL & 2.99\tablenotemark{c} & 0.01 & 0.41 & & $K_{\rm
  MEKAL} = 9.5 \times 10^{-3}$ & 1554 / 140 \\

North & Brem + PL + lines & 2.99\tablenotemark{c} & &
$0.36^{+0.01}_{-0.02}$ & $0.92^{+0.28}_{-0.19}$ & $\left\{
  \begin{array}{c} K_{\rm Brem} = \left( 3.1^{+0.1}_{-0.1} \right) \times
    10^{-3} \\ K_{\rm PL} = \left( 1.7^{+1.2}_{-0.2} \right) \times
    10^{-5} \end{array} \right\}$ & 106 / 92 \\

East & MEKAL & 2.99\tablenotemark{c} & 0.06 & 0.57 & & $K_{\rm MEKAL}
= 1.2 \times 10^{-3}$ & 328 / 108 \\

East & Brem + PL + lines & 2.99\tablenotemark{c} & &
$0.53^{+0.03}_{-0.04}$ & $-0.21^{+0.68}_{-0.77}$ & $\left\{
  \begin{array}{c} K_{\rm Brem} = \left( 3.9^{+0.3}_{-0.2} \right)
    \times 10^{-4} \\ K_{\rm PL} = \left( 1.6^{+3.9}_{-0.7} \right)
    \times 10^{-6} \end{array} \right\}$ & 88 / 73 \\

South & MEKAL & 2.99\tablenotemark{c} & 0.09 & 0.75 & & $K_{\rm
  MEKAL} = 1.5 \times 10^{-3}$ & 1228 / 156 \\

South & Brem + PL + lines & 2.99\tablenotemark{c} & &
$1.06^{+0.05}_{-0.05}$ & $-0.20^{+0.07}_{-0.06}$ & $\left\{
  \begin{array}{c} K_{\rm Brem} = \left( 3.3^{+0.1}_{-0.1} \right)
    \times 10^{-4} \\ K_{\rm PL} = \left( 4.7^{+1.7}_{-0.5} \right)
    \times 10^{-6} \end{array} \right\}$ & 144 / 118 \\

\enddata

\tablenotetext{a}{The model abbreviations are MEKAL = Mewe, Kaastra \&
  Liedahl thermal plasma, PL = power law, Brem = bremsstrahlung.}

\tablenotetext{b}{The model normalization are $K_{\rm MEKAL} =
  10^{-14} \int n_e n_H dV / (4 \pi ((1+z)D_A)^2)$, $K_{\rm Brem} =
  3.02 \times 10^{-15} \int n_e n_I dV / (4 \pi D^2)$ and $K_{\rm PL}
  = \phpcmsqps {\rm keV}^{-1}$ at 1 keV. $n_e$ is the electron density
  (\pcmcu), $n_H$ is the hydrogen density (\pcmcu), $n_I$ is the ion
  density (\pcmcu), $D$ is the distance to the source (cm) and $D_A$
  is the angular size distance to the source (cm).}

\tablenotetext{c}{Fixed parameter.}

\end{deluxetable}

\vfil\eject\clearpage
\begin{deluxetable}{cccc}
\tabletypesize{\footnotesize}
\tablewidth{0pt}
\tablecaption{X-ray Spectral Lines From the West Region
\label{tab:west_lines}}
\tablecolumns{4} \tablehead{\colhead{Energy} &
  \colhead{Line} & \colhead{Observed energy} &
  \colhead{$K$\tablenotemark{a}} \\ \colhead{[keV]}
  & \colhead{} & \colhead{[keV]} & \colhead{}} \startdata

0.37 & C {\sc vi} Ly $\alpha$ & $0.33^{+0.02}_{-0.01}$ & $\left(
  1.6^{+1.5}_{-0.6} \right) \times 10^{-5}$ \\

0.56--0.57 & O {\sc vii} triplet & $0.57^{+0.00}_{-0.01}$ &
$\left( 2.4^{+0.4}_{-0.3} \right) \times 10^{-5}$ \\

$\left\{ \begin{array}{c} 0.65 \\ 0.67 \end{array} \right.$ &
\begin{tabular}{c} O {\sc viii} Ly $\alpha$ \\ N {\sc vii} RRC
\end{tabular} $\left. \begin{array}{c} \\ \\ \end{array} \right\}$ &
$0.66^{+0.01}_{-0.00}$ & $\left( 3.0^{+0.4}_{-0.2} \right) \times
10^{-5}$ \\

0.74 & O {\sc vii} RRC \& Fe {\sc xvii} & $0.74^{+0.01}_{-0.01}$ &
$\left( 3.7^{+0.2}_{-0.4} \right) \times 10^{-5}$ \\

$\left\{ \begin{array}{c} 0.77 \\ 0.78 \\ 0.81 {\rm ~\&~} 0.82
  \end{array} \right.$ & \begin{tabular}{c} O {\sc viii} Ly $\beta$ \\
  Fe L {\sc xviii} \\ Fe L {\sc xvii} \end{tabular}
$\left. \begin{array}{c} \\ \\ \\
  \end{array} \right\}$ & $0.80^{+0.01}_{-0.01}$ & $\left(
  3.5^{+0.3}_{-0.3} \right) \times 10^{-5}$ \\

0.87 & O {\sc viii} RRC & $0.87^{+0.00}_{-0.01}$ & $\left(
  2.8^{+0.3}_{-0.3} \right) \times 10^{-5}$ \\

0.90 -- 0.92 & Ne {\sc ix} triplet & $0.92^{+0.01}_{-0.00}$ &
$\left( 2.7^{+0.2}_{-0.7} \right) \times 10^{-5}$ \\

1.02 & Ne {\sc x} Ly $\alpha$ & $1.01^{+0.01}_{-0.00}$ & $\left(
  2.2^{+0.2}_{-0.2} \right) \times 10^{-5}$ \\

1.07 & Ne {\sc ix} He $\beta$ & $1.11^{+0.00}_{-0.02}$ & $\left(
  5.4^{+1.1}_{-1.3} \right) \times 10^{-6}$ \\

1.23 & Na {\sc xi} Ly $\alpha$ & $1.23^{+0.01}_{-0.02}$ & $\left(
  4.4^{+0.9}_{-1.2} \right) \times 10^{-6}$ \\

1.33 -- 1.34 & Mg {\sc xi} triplet & $1.33^{+0.01}_{-0.01}$ & $\left(
  4.7^{+1.0}_{-0.9} \right) \times 10^{-6}$ \\

1.47 & Mg {\sc xii} Ly $\alpha$ & $1.44^{+0.03}_{-0.03}$ & $\left(
  1.3^{+0.4}_{-0.7} \right) \times 10^{-6}$ \\

1.84 -- 1.86 & Si {\sc xii} -- Si {\sc xiii} triplet &
$1.85^{+0.01}_{-0.01}$ & $\left( 3.8^{+0.9}_{-0.7} \right) \times 10^{-6}$ \\

2.31 -- 2.41 & S {\sc ii} -- S {\sc xiii} & $2.36^{+0.05}_{-0.04}$ &
$\left( 1.1^{+0.8}_{-0.6} \right) \times 10^{-6}$ \\

6.40 -- 6.97 & Fe {\sc ii} -- Fe {\sc xxvi} & 6.40\tablenotemark{b} &
$\left( 2.1^{+1.2}_{-1.1} \right) \times 10^{-6}$ \\

\enddata

\tablenotetext{a}{$K = {\rm total} \phpcmsqps$ in the line.}
\tablenotetext{b}{Fixed parameter.}

\end{deluxetable}

\vfil\eject\clearpage
\begin{deluxetable}{cccc}
\tabletypesize{\footnotesize}
\tablewidth{0pt}
\tablecaption{X-ray Spectral Lines From the North Region
\label{tab:north_lines}}
\tablecolumns{4} \tablehead{\colhead{Energy} &
  \colhead{Line} & \colhead{Observed energy} &
  \colhead{$K$\tablenotemark{a}} \\ \colhead{[keV]}
  & \colhead{} & \colhead{[keV]} & \colhead{}} \startdata

0.37 & C {\sc vi} Ly $\alpha$ & $0.38^{+0.00}_{-0.01}$ & $\left(
  1.2^{+0.1}_{-0.1} \right) \times 10^{-4}$ \\

0.43 & N {\sc vi} triplet & $0.43^{+0.01}_{-0.01}$ & $\left(
  5.3^{+1.0}_{-1.1} \right) \times 10^{-5}$ \\

0.52 & N {\sc vi} He $\gamma$ & $0.52^{+0.01}_{-0.00}$ & $\left(
  5.8^{+0.7}_{-0.8} \right) \times 10^{-5}$ \\

0.56 -- 0.57 & O {\sc vii} triplet & $0.57^{+0.01}_{-0.00}$ & $\left(
  1.0^{+0.1}_{-0.1} \right) \times 10^{-4}$ \\

0.67 & N {\sc vii} RRC \& O {\sc viii} Ly $\alpha$ &
$0.68^{+0.00}_{-0.01}$ & $\left( 4.9^{+0.5}_{-0.5} \right) \times
10^{-5}$ \\

0.74 & O {\sc vii} RRC \& Fe {\sc xvii} & $ 0.75^{+0.01}_{-0.00}$ &
$\left( 3.2^{+0.4}_{-0.4} \right) \times 10^{-5}$ \\

$\left\{ \begin{array}{c} 0.81 {\rm ~\&~} 0.82 \\ 0.82 \end{array} \right.$ &
\begin{tabular}{c} Fe {\sc xvii} \\ O {\sc viii} Ly $\gamma$
\end{tabular} $\left. \begin{array}{c} \\ \\ \end{array} \right\}$ &
$0.83^{+0.01}_{-0.01}$ & $\left( 3.5^{+0.4}_{-0.3} \right) \times
10^{-5}$ \\

0.90 -- 0.92 &  Ne {\sc ix} triplet & $0.90^{+0.00}_{-0.01}$ & $\left(
  5.0^{+0.3}_{-0.4} \right) \times 10^{-5}$ \\

$\sim 0.95$ & Fe L {\sc xx} & $0.95^{+0.01}_{-0.01}$ &
$\left( 3.1^{+0.6}_{-0.6} \right) \times 10^{-5}$ \\

1.02 & Ne {\sc x} Ly $\alpha$ & $1.03^{+0.00}_{-0.00}$ & $\left(
  1.9^{+0.3}_{-0.1} \right) \times 10^{-5}$ \\


1.15 & Ne {\sc ix} He $\gamma$ & $1.14^{+0.01}_{-0.01}$ & $\left(
  7.5^{+1.6}_{-1.8} \right) \times 10^{-6}$ \\

1.23 & Na {\sc xi} Ly $\alpha$ & $1.22^{+0.01}_{-0.01}$ & $\left(
  4.6^{+1.7}_{-1.2} \right) \times 10^{-6}$ \\

1.33 -- 1.34 & Mg {\sc xi} triplet & $1.34^{+0.01}_{-0.01}$ & $\left(
  7.8^{+1.4}_{-1.3} \right) \times 10^{-6}$ \\

1.47 & Mg {\sc xii} Ly $\alpha$ & $1.47^{+0.02}_{-0.03}$ & $\left(
  1.6^{+0.9}_{-0.9} \right) \times 10^{-6}$ \\

1.74 -- 1.77 & Si {\sc ii} -- Si {\sc viii} & $1.76^{+0.01}_{-0.02}$ &
$\left( 3.4^{+1.0}_{-1.1} \right) \times 10^{-6}$ \\

1.82 -- 1.86 & Si {\sc xi} -- Si {\sc xiii} triplet &
$1.83^{+0.01}_{-0.01}$ & $\left( 5.0^{+0.8}_{-1.0} \right) \times
10^{-6}$ \\

2.00 & Si {\sc xiv} & $1.97^{+0.02}_{-0.03}$ & $\left(
  1.8^{+0.7}_{-0.7} \right) \times 10^{-6}$ \\

2.31 -- 2.37 & S {\sc ii} -- S {\sc xi} & $2.35^{+0.02}_{-0.02}$ &
$\left( 2.9^{+0.9}_{-0.9} \right) \times 10^{-6}$ \\

2.46 & S {\sc xv} triplet & $2.49^{+0.02}_{-0.03}$ & $\left(
  1.4^{+0.7}_{-0.7} \right) \times 10^{-6}$ \\

3.01 -- 3.14 & Ar {\sc xii} -- Ar {\sc xvii} triplet &
$3.06^{+0.05}_{-0.04}$ & $\left( 1.0^{+0.7}_{-0.6} \right) \times
10^{-6}$ \\

3.84 -- 4.11 & Ca {\sc xvii} -- Ca {\sc xx} & $3.93^{+0.04}_{-0.08}$ &
$\left( 8.1^{+5.8}_{-5.7} \right) \times 10^{-7}$ \\

6.40 -- 6.97 & Fe {\sc ii} -- Fe {\sc xxvi} & $6.32^{+0.05}_{-0.03}$ &
$\left( 2.6^{+1.4}_{-1.4} \right) \times 10^{-6}$ \\

\enddata

\tablenotetext{a}{$K = {\rm total} \phpcmsqps$ in the line.}

\end{deluxetable}

\vfil\eject\clearpage
\begin{deluxetable}{cccc}
\tabletypesize{\footnotesize}
\tablewidth{0pt}
\tablecaption{X-ray Spectral Lines From the East Region
\label{tab:east_lines}}
\tablecolumns{4} \tablehead{\colhead{Energy} &
  \colhead{Line} & \colhead{Observed energy} &
  \colhead{$K$\tablenotemark{a}} \\ \colhead{[keV]}
  & \colhead{} & \colhead{[keV]} & \colhead{}} \startdata


0.43 & N {\sc vi} triplet & $0.41^{+0.01}_{-0.02}$ & $\left(
  7.9^{+4.9}_{-4.6} \right) \times 10^{-6}$ \\

0.50 & N {\sc vii} & $0.52^{+0.02}_{-0.01}$ & $\left(
  8.4^{+3.0}_{-4.0} \right) \times 10^{-6}$ \\

0.59 & N {\sc vii} Ly $\alpha$ & $0.59^{+0.01}_{-0.01}$ & $\left(
  1.9^{+0.3}_{-0.5} \right) \times 10^{-5}$ \\

0.65 & N {\sc vii} RRC & $0.67^{+0.01}_{-0.01}$ & $\left(
  2.0^{+0.2}_{-0.4} \right) \times 10^{-5}$ \\

0.73 & Fe {\sc xvii} & $0.73^{+0.01}_{-0.01}$ & $\left(
  2.0^{+0.4}_{-0.3} \right) \times 10^{-5}$ \\

$\left\{ \begin{array}{c} 0.81 {\rm ~\&~} 0.82 \\ 0.82 \end{array}
\right.$ & \begin{tabular}{c} Fe {\sc xvii} \\ O {\sc viii} Ly
  $\delta$ \end{tabular} $\left. \begin{array}{c} \\ \\ \end{array}
\right\}$ & $0.80^{+0.02}_{-0.02}$ & $\left( 2.1^{+0.2}_{-0.4} \right)
\times 10^{-5}$ \\

0.87 & O {\sc viii} RRC & $0.85^{+0.02}_{-0.00}$ & $\left(
  1.7^{+0.2}_{-0.5} \right) \times 10^{-5}$ \\ 

0.90 -- 0.92 & Ne {\sc ix} triplet & $0.91^{+0.01}_{-0.00}$ &
$\left( 2.0^{+0.3}_{-0.5} \right) \times 10^{-5}$ \\

1.02 & Ne {\sc x} & $1.01^{+0.00}_{-0.01}$ & $\left( 1.2^{+0.1}_{-0.2}
\right) \times 10^{-5}$ \\

1.07 & Ne {\sc ix} He $\beta$ & $1.08^{+0.01}_{-0.02}$ & $\left(
  3.8^{+1.1}_{-1.1} \right) \times 10^{-6}$ \\

1.15 & Ne {\sc ix} He $\gamma$ & $1.15^{+0.02}_{-0.02}$ & $\left(
  2.6^{+0.9}_{-0.7} \right) \times 10^{-6}$ \\

1.30 -- 1.35 & Mg {\sc viii} -- Mg {\sc xi} triplet &
$1.32^{+0.01}_{-0.02}$ & $\left( 3.1^{+0.9}_{-1.1} \right) \times
10^{-6}$ \\

1.35 & Mg {\sc xi} triplet & $1.37^{+0.02}_{-0.03}$ & $\left(
  1.1^{+0.7}_{-0.7} \right) \times 10^{-6}$ \\

1.48 & Mg {\sc xii} Ly $\alpha$ & $1.47^{+0.02}_{-0.03}$ & $\left(
  7.8^{+6.0}_{-5.7} \right) \times 10^{-7}$ \\

1.78 -- 1.84 & Si {\sc ix} -- Si {\sc xii} & $1.81^{+0.01}_{-0.02}$ &
$\left( 2.0^{+0.6}_{-0.6} \right) \times 10^{-6}$ \\

1.87 -- 2.00 & Si {\sc xiii} -- Si {\sc xiv} & $1.93^{+0.03}_{-0.03}$ &
$\left( 1.0^{+0.5}_{-0.5} \right) \times 10^{-6}$ \\

2.35 -- 2.46 & S {\sc x} -- S {\sc xv} & $2.38^{+0.06}_{-0.03}$ &
$\left( 1.2^{+0.6}_{-0.6} \right) \times 10^{-6}$ \\

\enddata

\tablenotetext{a}{$K = {\rm total} \phpcmsqps$ in the line.}

\end{deluxetable}

\vfil\eject\clearpage
\begin{deluxetable}{cccc}
\tabletypesize{\footnotesize}
\tablewidth{0pt}
\tablecaption{X-ray Spectral Lines From the South Region
\label{tab:south_lines}}
\tablecolumns{4} \tablehead{\colhead{Energy} &
  \colhead{Line} & \colhead{Observed energy} &
  \colhead{$K$\tablenotemark{a}} \\ \colhead{[keV]}
  & \colhead{} & \colhead{[keV]} & \colhead{}} \startdata

0.37 & C {\sc vi} & $0.37^{+0.01}_{-0.01}$ & $\left(
  2.2^{+0.8}_{-0.8} \right) \times 10^{-5}$ \\

0.43 & N {\sc vi} triplet & $0.43^{+0.01}_{-0.01}$ & $\left(
  2.0^{+0.5}_{-0.7} \right) \times 10^{-5}$ \\

0.50 & N {\sc vii} & $0.50^{+0.01}_{-0.01}$ & $\left(
  1.3^{+0.4}_{-0.5} \right) \times 10^{-5}$ \\

0.56 & O {\sc vii} triplet & $0.56^{+0.01}_{-0.00}$ & $\left(
  4.0^{+0.6}_{-1.2} \right) \times 10^{-5}$ \\

0.67 & N {\sc vii} RRC \& O {\sc viii} Ly $\alpha$ &
$0.67^{+0.00}_{-0.01}$ & $\left( 3.7^{+0.3}_{-1.9} \right) \times
10^{-5}$ \\

0.72 & O {\sc vii} RRC \& Fe {\sc xvii} & $0.74^{+0.00}_{-0.01}$ &
$\left( 4.3^{+0.4}_{-1.3} \right) \times 10^{-5}$ \\

$\left\{ \begin{array}{c} 0.79 - 0.82 \\ 0.77 - 0.82 \end{array}
\right.$ & \begin{tabular}{c} Fe {\sc xvii} \\ O {\sc viii} Ly
  $\beta$, $\gamma$ \end{tabular} $\left. \begin{array}{c} \\ \\
  \end{array} \right\}$ & $0.80^{+0.07}_{-0.04}$ & $\left(
  3.3^{+0.5}_{-1.6} \right) \times 10^{-5}$ \\

0.79 -- 0.82 & O {\sc viii} Ly $\alpha$ \& Fe {\sc xvii} &
$0.84^{+0.02}_{-0.04}$ & $\left( 3.3^{+0.5}_{-1.0} \right) \times
10^{-5}$ \\

0.90 -- 0.92 & Ne {\sc ix} triplet & $0.90^{+0.00}_{-0.01}$ &
$\left( 4.0^{+0.3}_{-2.0} \right) \times 10^{-5}$ \\

0.97 & Fe {\sc xx} & $0.96^{+0.01}_{-0.00}$ & $\left(
  1.9^{+0.2}_{-0.3} \right) \times 10^{-5}$ \\

1.02 & Ne {\sc x} Ly $\alpha$ & $1.03^{+0.01}_{-0.00}$ & $\left(
  2.1^{+0.2}_{-0.2} \right) \times 10^{-5}$ \\

1.15 & Ne {\sc ix} He $\gamma$ & $1.12^{+0.01}_{-0.01}$ & $\left(
  7.1^{+1.5}_{-1.5} \right) \times 10^{-6}$ \\

1.21 & Ne {\sc x} Ly $\beta$ & $1.21^{+0.01}_{-0.01}$ & $\left(
  8.1^{+1.3}_{-1.5} \right) \times 10^{-6}$ \\

1.35 & Mg {\sc xi} triplet & $1.33^{+0.00}_{-0.01}$ & $\left(
  1.0^{+0.1}_{-0.1} \right) \times 10^{-5}$ \\

1.47 & Mg {\sc xii} Ly $\alpha$ & $1.45^{+0.02}_{-0.02}$ & $\left(
  2.6^{+0.9}_{-1.0} \right) \times 10^{-6}$ \\

1.74 -- 1.77 & Si {\sc ii} -- Si {\sc viii} triplet &
$1.74^{+0.02}_{-0.01}$ & $\left( 3.0^{+1.0}_{-0.9} \right) \times
10^{-6}$ \\

1.82 -- 1.86 & Si {\sc xi} -- Si {\sc xiii} triplet &
$1.83^{+0.01}_{-0.01}$ & $\left( 5.6^{+1.0}_{-1.1} \right) \times
10^{-6}$ \\

2.39 -- 2.46 & S {\sc xii} -- S {\sc xv} triplet &
$2.42^{+0.02}_{-0.03}$ & $\left( 1.9^{+1.8}_{-0.9} \right) \times
10^{-6}$ \\

6.40 & Fe {\sc ii} & 6.40\tablenotemark{b} &
$\left( 5.1^{+1.7}_{-1.6} \right) \times 10^{-6}$ \\

\enddata

\tablenotetext{a}{$K = {\rm total} \phpcmsqps$ in the line.}
\tablenotetext{b}{Fixed parameter.}

\end{deluxetable}

\vfil\eject\clearpage
\begin{deluxetable}{cccc}
\tabletypesize{\footnotesize}
\tablewidth{0pt}
\tablecaption{Fluxes and Luminosities of the Larger Scale Extended
  X-ray Emission \label{tab:extended_flux}}
\tablecolumns{4} \tablehead{\colhead{Region} & \colhead{Energy band} &
  \colhead{Unabsorbed flux} & \colhead{Unabsorbed luminosity} \\ 
  \colhead{} & \colhead{[keV]} & \colhead{[$\ergpcmsqps$]} &
  \colhead{[$\ergps$]}} \startdata

West & 0.5 -- 2.0 & $5.8 \times 10^{-13}$ & $3.6 \times 10^{40}$ \\

West & 2.0 -- 10.0 & $2.3 \times 10^{-13}$ & $1.4 \times 10^{40}$ \\

North & 0.5 -- 2.0 & $1.4 \times 10^{-12}$ & $8.7 \times 10^{40}$ \\

North & 2.0 -- 10.0 & $3.3 \times 10^{-13}$ & $2.0 \times 10^{40}$ \\

East & 0.5 -- 2.0 & $4.1 \times 10^{-13}$ & $2.5 \times 10^{40}$ \\

East & 2.0 -- 10.0 & $2.0 \times 10^{-13}$ & $1.2 \times 10^{40}$ \\

South & 0.5 -- 2.0 & $7.8 \times 10^{-13}$ & $4.9 \times 10^{40}$ \\

South & 2.0 -- 10.0 & $6.6 \times 10^{-13}$ & $4.0 \times 10^{40}$ \\

\enddata
\end{deluxetable}

\vfil\eject\clearpage

\vfil\eject\clearpage

\vfil\eject\clearpage

\vfil\eject\clearpage

\vfil\eject\clearpage

\vfil\eject\clearpage

\vfil\eject\clearpage

\vfil\eject\clearpage

\vfil\eject\clearpage

\vfil\eject\clearpage

\vfil\eject\clearpage

\vfil\eject\clearpage
\begin{figure}
\centerline{\psfig{figure=f12.ps,width=0.65\textwidth,angle=270}}
\caption{X-ray spectrum of the South region (defined in
  section \ref{sec:reduction_large}) of NGC~1068 extracted from the
  3.2 s frame-time data. The upper panel shows the data points with
  error bars (crosses), with the model folded through the instrument
  response (uppermost solid line passing through the data points). The
  individual components of the model are plotted below this.  The
  lower panel shows the $\chi$ residuals to this fit.  The parameters
  of this fit are listed in tables \ref{tab:extended_cont} and
  \ref{tab:south_lines}. Note that the calibration is uncertain below
  0.50 keV and degrades rapidly below 0.45 keV. The four large
  diagonal crosses with horizontal bars represent estimates of the
  contributions in the indicated energy bands from the bright, inner
  ($r < 5 \arcsec$) emission which has been scattered by the telescope
  PSF (see section \ref{sec:psf}).
\label{fig:south}}
\end{figure}

\vfil\eject\clearpage

\vfil\eject\clearpage

\vfil\eject\clearpage

\vfil\eject\clearpage

\vfil\eject\clearpage

\vfil\eject\clearpage

\vfil\eject\clearpage


\begin{thebibliography}{}
  
\bibitem[Arnaud \& Raymond (1992)]{ar92} Arnaud, M. \& Raymond, J. 1992,
  ApJ, 398, 394
  
\bibitem[Arnaud \& Rothenflug (1985)]{ar85} Arnaud, M. \& Rothenflug, M.
  1985, A\&AS, 60, 425
  
\bibitem[Baldwin, Wilson \& Whittle (1987)]{bww87} Baldwin, J. A.,
  Wilson, A. S. \& Whittle, M. 1987, ApJ, 319, 84
  
  
\bibitem[Bock \etal (2000)]{betal00} Bock, J. J., Neugebauer, G.,
  Matthews, K., Soifer, B. T., Becklin, E. E., Ressler, M., Marsh, K.,
  Werner, M. W., Egami, E. \& Blandford, R. 2000, AJ, 120, 2904
  
\bibitem[Braatz \etal (1993)]{betal93} Braatz, J. A., Wilson, A. S.,
  Gezari, D. Y., Varosi, F. \& Beichman, C. A. 1993, ApJ, 409, L5
  
\bibitem[Capetti, Axon \& Macchetto (1997)]{cetal97} Capetti, A.,
  Axon, D. J. \& Macchetto, F. D. 1997, ApJ, 487, 560
  
\bibitem[Capetti \etal (1995)]{cetal95} Capetti, A., Macchetto, F.,
  Axon, D. J., Sparks, W. B. \& Boksenberg, A. 1995, ApJ, 452. L87

\bibitem[Capetti, Machetto \& Lattanzi (1997)]{cml97} Capetti, A.,
  Machetto, F. D. \& Lattanzi, M. G. 1997, ApJ, 476, L67
  
\bibitem[Claussen \& Lo (1986)]{cl96} Claussen, M. J. \& Lo, K.-Y.
  1986, ApJ, 308, 592
  
\bibitem[Davis (2001)]{d01} Davis, J. E. 2001, ApJ, submitted

\bibitem[Dressel \etal (1997)]{detal97} Dressel, L. L., Tsvetanov, Z.
  I., Kriss, G. A. \& Ford, H. C. 1997, Ap\&SS, 248, 85

\bibitem[Elvis \& Lawrence (1988)]{el88} Elvis, M. \& Lawrence,
  A. 1988, ApJ, 331, 161

\bibitem[Evans \etal (1991)]{eetal91} Evans, I. N., Ford, H. C.,
  Kinney, A. L., Antonucci, R. R. J., Armus, L. \& Caganoff, S. 1991,
  ApJ, 369, L27
  
  
\bibitem[Gallimore, Baum \& O'Dea (1997)]{gbo96} Gallimore, J. F.,
  Baum, S. A. \& O'Dea, C. P. 1996, ApJ, 464, 198
  
\bibitem[Gallimore, Baum \& O'Dea (1997)]{gbo97} Gallimore, J. F.,
  Baum, S.  A.  \& O'Dea, C. P. 1997, Nature, 388, 852

\bibitem[Gallimore \etal (1994)]{getal94} Gallimore, J. F., Baum, S.
  A., O'Dea, C. P., Brinks, E. \& Pedlar, A. 1994, ApJ, 422, L13
  
\bibitem[Gallimore \etal (1996)]{g1etal96} Gallimore, J. F., Baum, S.
  A., O'Dea, C. P. \& Pedlar, A. 1996, ApJ, 458, 136

\bibitem[Garmire \etal (2000)]{getal00} Garmire, G. P., \etal 2000,
  ApJS, submitted
  
\bibitem[Greenhill \etal (1996)]{g2etal96} Greenhill, L. J., Gwinn, C.
  R., Antonucci, R. R. J. \& Barvainis, R.  1996, ApJ, 472, L21

\bibitem[Guainazzi \etal (1999)]{getal99} Guainazzi, M., \etal 1999,
  MNRAS, 310, 10
  
  
  
\bibitem[Iwasawa, Fabian \& Matt (1997)]{ifm97} Iwasawa, K., Fabian, A.
  C. \& Matt, G. 1997, MNRAS, 289, 443
  
\bibitem[Kaastra (1992)]{k92} Kaastra, J. S. 1992, An X-Ray Spectral Code
  for Optically Thin Plasmas (Internal SRON-Leiden Report, updated
  version 2.0)
  
\bibitem[Kishimoto (1999)]{k99} Kishimoto, M. 1999, ApJ, 518, 676

\bibitem[Koyama \etal (1989)]{ketal89} Koyama, K., Inoue, H., Tanaka,
  Y., Awaki, H., Takano, S., Ohashi, T. \& Matsuoka, M. 1989, PASJ,
  41, 731

\bibitem[Krolik \& Kallman (1987)]{kk87} Krolik, J. H. \& Kallman,
  T. R. 1987, ApJ, 320, L5

\bibitem[Liedahl \etal (1995)]{letal95} Liedahl, D.A., Osterheld, A.L.
  \& Goldstein, W. H. 1995, ApJ, 438, L115
  
\bibitem[Machetto \etal (1994)]{metal94} Macchetto, F. D., Capetti,
  A., Sparks, W. B., Axon, D. J. \& Boksenberg, A. 1994, ApJ, 435,
  L15


\bibitem[Marshall \etal (1993)]{metal93} Marshall, F. E., \etal 1993,
  ApJ, 405, 168
  
\bibitem[Matt \etal (1997)]{metal97} Matt, G., \etal 1997, A\&A, 325,
  L13
  
\bibitem[Matt \etal (2000)]{metal00} Matt, G., Fabian, A. C.,
  Guainazzi, M., Iwasawa, K., Bassani, L. \& Malaguti, G. 2000, MNRAS,
  318, 173
  
\bibitem[Mewe \etal (1985)]{metal85} Mewe, R., Gronenschild, E. H. B.
  M.  \& van den Oord, G.H.J. 1985, A\&AS, 62, 179
    
\bibitem[Mewe \etal (1986)]{metal86} Mewe, R., Lemen, J.R. \& van den
  Oord, G. H. J. 1986, A\&AS, 65, 511

\bibitem[Monier \& Halpern (1987)]{mh87} Monier, R. \& Halpern,
  J. P. 1987, ApJ, 315, L17

\bibitem[Murphy \etal (1996)]{m1etal96} Murphy, M. M., Lockman, F. J.,
  Laor, A., \& Elvis, M. 1996, ApJS, 105, 369
    
\bibitem[Muxlow \etal (1996)]{m2etal96} Muxlow, T. W. B., Pedlar, A.,
  Holloway, A. J., Gallimore, J. F.  \& Anotnucci, R. R. J.  1996,
  MNRAS, 278, 854
  
\bibitem[Netzer \& Turner (1997)]{nt97} Netzer, H. \& Turner, T. J.
  1997, ApJ, 488, 694
  
\bibitem[Paerels \etal (2000)]{petal00} Paerels, F., \etal 2000, Bull.
  A.A.S., 32, 1181
  
\bibitem[Pogge \& De Robertis (1993)]{pdr93} Pogge, R. W. \& De
  Robertis, M. M. 1993, ApJ, 404, 563
  

\bibitem[Roy \etal (1998)]{retal98} Roy, A. L., Colbert, E. J. M.,
  Wilson, A. S. \& Ulvestad, J. S. 1998, ApJ, 504, 147
  
\bibitem[Thatte \etal (1997)]{tetal97} Thatte, N., Quirrenbach, A.,
  Genzel, R., Maiolino, R. \& Tecza, M. 1997, ApJ, 490, 238
  
\bibitem[Ueno \etal (1994)]{uetal94} Ueno, S., Mushotzsky, R. F.,
  Koyama, K., Iwasawa, K., Awaki, H. \& Hayashi, I. 1994, PASP, 46,
  L71

\bibitem[Ulvestad, Neff \& Wilson (1987)]{unw87} Ulvestad, J. S.,
  Neff, S. G. \& Wilson, A. S. 1987, AJ, 92, 22
  
\bibitem[de Vaucouleurs \etal (1991)]{vetal91} de Vaucouleurs, G., de
  Vaucouleurs, A., Corwin, H. G., Buta, R. J., Paturel, G. \&
  Fouqu\'e, P. 1991, Third Reference Catalogue of Bright Galaxies,
  Springer-Verlag

  
\bibitem[Weinberger, Neugebauer \& Matthews (1999)]{wnm99} Weinberger,
  A. J., Neugebauer, G. \& Matthews, K. 1999, AJ, 117, 2748
  
\bibitem[Wilson \& Elvis (1997)]{we97} Wilson, A. S. \& Elvis, M. 1997
  in Proceedings of the Schloss Ringberg NGC 1068 Workshop, eds J.  F.
  Gallimore \& L. R. Tacconi, Ap\&SS, 248, 167

\bibitem[Wilson \etal (1992)]{wetal92} Wilson, A. S., Elvis, M.,
  Lawrence, A. \& Bland-Hawthorn, J. 1992, ApJ, 391, L75

\bibitem[Wilson \& Ulvestad (1983)]{wu83} Wilson, A. S. \& Ulvestad,
  J. S. 1983, ApJ, 275, 8 (WU)

\bibitem[Wilson \& Ulvestad (1987)]{wu87} Wilson, A. S. \& Ulvestad,
  J. S. 1987, ApJ, 319, 105
  
\bibitem[Wilson, Ward \& Haniff (1988)]{wwh88} Wilson, A. S., Ward, M.
  J. \& Haniff, C. A. 1988, ApJ, 334, 121
  
\bibitem[Wilson \& Willis (1980)]{ww80} Wilson, A. S. \& Willis, A. G.
  1980, ApJ, 240, 429

\end{thebibliography}
\end{document}